\documentclass[%
 aip,
 sd,%
 amsmath,amssymb,
 reprint,%
]{revtex4-1}

\usepackage{graphicx}
\usepackage{epstopdf}
\usepackage{dcolumn}
\usepackage{bm}
\usepackage{amsthm}

\theoremstyle{plain}
\newtheorem{theorem}{Theorem}

\newtheorem{corollary}[theorem]{Corollary}
\newtheorem{proposition}[theorem]{Proposition}

\theoremstyle{definition}
\newtheorem{definition}{Definition}

\theoremstyle{remark}
\newtheorem{remark}{Remark}

\begin{document}

\preprint{AIP/123-QED}

\title[]{Phase holonomy underlies puzzling temporal patterns in Kuramoto models with two sub-populations}

\author{Aladin Crnki\' c}%
\email{aladin.crnkic@unbi.ba}
\affiliation{
Faculty of Technical Engineering, University of Biha\' c, I. Ljubijanki\' ca, bb., 77000 Biha\' c, Bosnia and Herzegovina.
}%
\author{Vladimir Ja\' cimovi\' c}
\email{vladimirj@ucg.ac.me}
\affiliation{
Faculty of Natural Sciences and Mathematics, University of Montenegro, Cetinjski put, bb., 81000 Podgorica, Montenegro.
}%

\date{\today}

\begin{abstract}
We present a geometric investigation of curious dynamical behaviors previously reported in Kuramoto models with two sub-populations. Our study demonstrates that chimeras and traveling waves in such models are associated with the birth of geometric phase. Although manifestations of geometric phase are frequent in various fields of Physics, this is the first time (to our best knowledge) that such a phenomenon is exposed in ensembles of Kuramoto oscillators or, more broadly, in complex systems.
\end{abstract}

\maketitle

\begin{quotation}
Kuramoto models display surprising spatio-temporal patterns even in seemingly simple setups. Two notable examples of this kind are the chimera state \cite{AMSW} and the traveling wave in the conformists-contrarians model \cite{HS2}. The existence of these stable equilibrium states in Kuramoto models with two sub-populations have been shown analytically using the Watanabe-Strogatz reduction \cite{WS} and group-theoretic approach \cite{MMS}.

At the first glance, it looks like these equilibria are stationary up to rotations: both the chimera and the traveling wave undergo a cyclic evolution, preserving their shape and coherence.

We demonstrate that this stationarity is deceptive. Although densities rotate, motions of individual oscillators are more complicated. This observation unveils impact of a "hidden variable": the ensemble exhibits a phase shift which is not a simple rotation. This hidden variable can be regarded as a novel manifestation of the geometric phase. Indeed, the phase shift arises from the geometric phenomenon of (an)holonomy from parallel translations for a connection on circle bundle with the holonomy group $SU(1, 1)$. This precisely fits into the mathematical framework that underlies the notion of geometric phase in Physics.

This unveils a geometric subtlety associated with intriguing temporal patterns in Kuramoto models with two sub-populations. Finally, we extend our analysis to recently reported traveling waves in conformists-contrarians models on spheres \cite{CJN}. We demonstrate that the emergence of these higher-dimensional traveling waves is associated with non-Abelian geometric phases which belong to the special orthogonal groups $SO(d)$.
\end{quotation}

\section{Introduction}
\label{sec:1}
Ensembles of Kuramoto oscillators display various intriguing spatio-temporal patterns. This has been observed in numerical studies soon after Kuramoto introduced his model \cite{Kuramoto}. The foundation for theoretical investigations of these patterns and related phenomena has been laid by Watanabe and Strogatz \cite{WS}. Their result, now commonly known as WS reduction, facilitates mathematical analysis of simple networks of Kuramoto oscillators by unveiling hidden symmetries and inferring low-dimensional dynamics on invariant submanifolds.

Later findings led to deeper understanding of symmetries and constants of motion in Kuramoto models \cite{MMS}. Analytic results have been obtained not only for ensembles with identical intrinsic frequencies, but also for heterogeneous ensembles (where frequencies are sampled from certain prescribed probability distributions) \cite{OA}.


Overall, group-theoretic and geometric investigations of spatio-temporal patterns in Kuramoto networks evolved into a very intriguing research direction. In our point of view, three aspects are of particular significance. First, connections between collective motions of Kuramoto oscillators and some classical mathematical theories, such as complex analysis, potential theory, hyperbolic geometry \cite{MMS,CEM,LMS}. Studies of Kuramoto models on spheres (and other manifolds) require a demanding and less investigated mathematical apparatus, thus presenting new challenges to researchers. Second, the group-theoretic approach enables rigorous analysis of some counter-intuitive dynamical behaviors in simple networks. Third, it has a potential to unveil unexpected relations with some other fields of Physics, where symmetries play a central role.

The present paper is a continuation of investigations initiated by the authors \cite{CJ-IJMP-B,JC-PhysicaD}. We focus on two well-known analytically solvable models that exhibit puzzling dynamical phenomena. The first phenomenon is the chimera state, reported by Abrams et al.\cite{AMSW} The second is traveling wave (TW) in the conformists-contrarians (conf-contr) model, studied by Hong and Strogatz \cite{HS2}. Finally, we briefly address traveling waves arising in conf-contr models on spheres, recently reported by the authors \cite{CJN}.

We demonstrate that these counter-intuitive dynamical behaviors are associated with a geometric subtlety. A closer look unveils that motions of traveling waves (including the chimera, since it also appears in the form of TW) are deceptively simple, and in order to understand these phenomena, one needs to investigate dynamics on the full space. Configuration space for TW's is a non-principal circle bundle acted on by group $PSU(1,1) = SU(1,1) / \pm I$. Parallel translation for a connection on this bundle gives rise to the phase shift on the circle that is not a simple rotation. From the point of view of theoretical physics, this phase shift can be regarded as a particular manifestation of the phenomenon named {\it geometric phase}.

The notion of geometric phase refers to various effects observed in different fields of Physics. The best known manifestations are in quantum physics, including the famous Berry phase \cite{Berry1} and the Aharonov-Anandan phase \cite{AA}.

However, geometric phases are widespread in classical physics as well. Related phenomena have been reported long before Berry's work, we refer to a brief survey on prehistory of his phase \cite{Berry2}. One example of the classical mechanical system that exhibits geometric phase is Foucault's pendulum \cite{vBx2}.

Although seemingly unrelated, all manifestations of the geometric phase can be explained through geometric concept of holonomy (also called anholonomy) resulting from the parallel transport. This mathematical point of view on geometric phases has been exposed \cite{Simon} immediately after the seminal Berry's paper \cite{Berry1}.

In the next Section we briefly explain two models and their puzzling equilibrium states that are explored here. We also present some preliminary simulation results, which indicate that these collective behaviors are more complicated than they appear at the first glance. In Section \ref{sec:3} we reduce the dynamics to the group of M\" obius transformations and its orbits.
Sections \ref{sec:4} and \ref{sec:5} present a numerical and analytic study of a geometric subtlety underlying the emergence of TW's in (\ref{chimera}) and (\ref{conf-contr}). In Section \ref{sec:6} we expose the effect of phase holonomy even in the simplest ensemble of identical, globally coupled oscillators. Section \ref{sec:7} contains a brief discussion on analogies with cyclic evolution on the invariant manifold of coherent states in quantum physics. In this analogy, the Poisson manifold corresponds to the manifold of coherent states. Section \ref{sec:8} deals with TW's that arise in the conf-contr models on spheres. It is shown that TW's on spheres are associated with $SO(d)$ non-Abelian geometric phases. Finally, Section \ref{sec:9} contains several concluding remarks.
\begin{figure}[t]
\centering
  \begin{tabular}{@{}c@{}}
    \includegraphics[width=.4\textwidth]{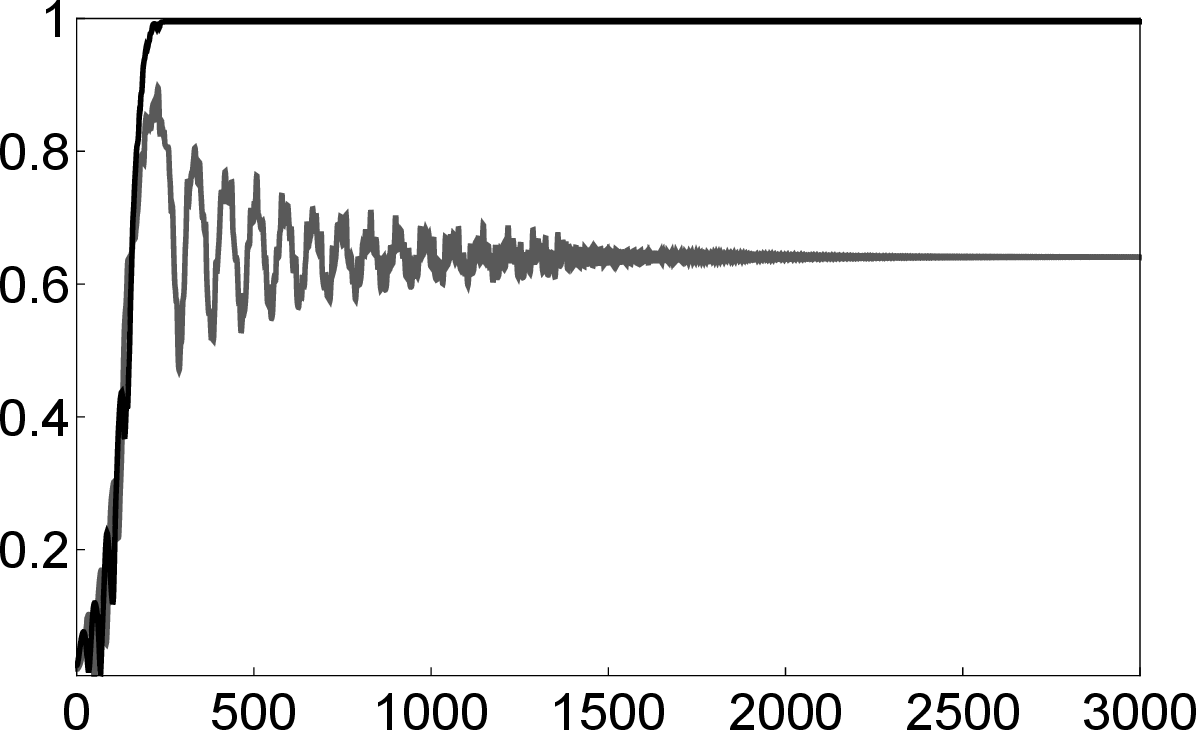}
  \end{tabular}
  \caption{\label{fig:1} Evolution of real order parameters of the two sub-populations in (\ref{chimera}) as the system evolves towards the stable chimera state. It shows that $r_A \to 1$, while $r_B \to const \approx 0.63.$ Obtained by solving (\ref{chimera}) for $2N = 500$ oscillators and parameter values $\mu = 0.623$, $\nu = 0.377$, $\beta = \frac{\pi}{2}-0.1$.}
\end{figure}

\section{Models and puzzles}
\label{sec:2}
\subsection{Stable chimera in the solvable chimera model}
\label{sec:2_1}
We start with the solvable chimera model, introduced and analyzed by Abrams et al. \cite{AMSW} This model describes an ensemble consisting of two sub-populations, denoted by $A$ and $B$:
\begin{equation}
\label{chimera}
\frac{d \varphi_j^l}{dt} = \omega + \sum \limits_{k=A,B} \frac{K_{kl}}{N} \sum \limits_{i=1}^{N} \sin(\varphi_i^k - \varphi_j^l - \beta),
\end{equation}
where $l=A,B$ and $j=\overline{1,N}$. Total number of oscillators is $2N$ ($N$ in each sub-population). All oscillators have equal intrinsic frequencies, denoted by $\omega \in {\mathbb R}$, and the coupling includes a global phase shift $\beta$. Furthermore, $K_{AA}$ and $K_{BB}$ are coupling strengths within each sub-population, while $K_{AB}$ and $K_{BA}$ are coupling strengths between pairs of oscillators that belong to different sub-populations. We assume that $K_{AA} = K_{BB} = \mu$, $K_{AB} = K_{BA} = \nu$ and $\mu > \nu$. This means that all pairwise couplings are symmetric, the coupling strengths within each sub-population are the same, but pairwise couplings between oscillators from different sub-populations are weaker than those between oscillators belonging to the same sub-population.

It has been shown that (\ref{chimera}) admits a peculiar {\it chimera state} \cite{AMSW}, a stable equilibrium in which one sub-population is fully synchronized, while the second sub-population remains only partially synchronized with constant real order parameter, see Figure \ref{fig:1}. Since the model is symmetric w.r. to permutation of sub-populations, it admits two symmetric chimera states; one in which $A$ is synchronized and $B$ desynchronized, and the second vice versa. In addition, both sub-populations travel along the circle, preserving the constant angle between them (this equilibrium value of the angle depends on parameters $\mu-\nu$ and $\beta$). This state has been named {\it stable chimera} \cite{AMSW}.
\begin{figure*}[t]
\centering
  \begin{tabular}{@{}cc@{}}
    \includegraphics[width=.4\textwidth]{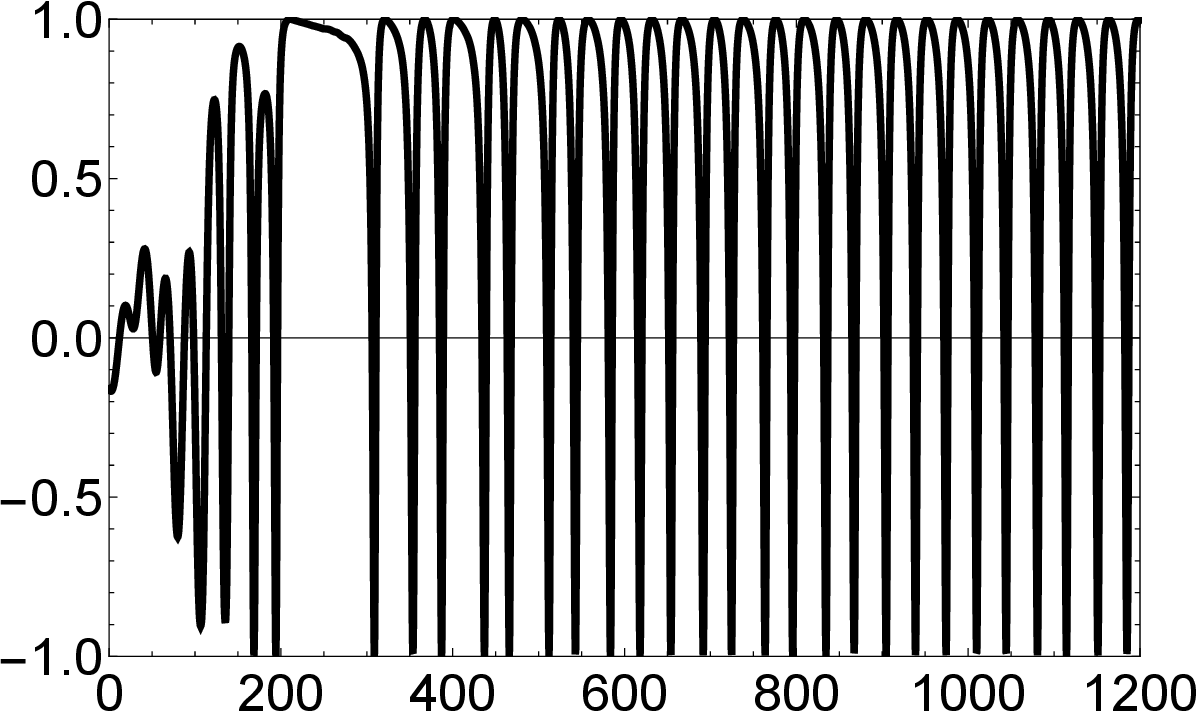} &
    \includegraphics[width=.4\textwidth]{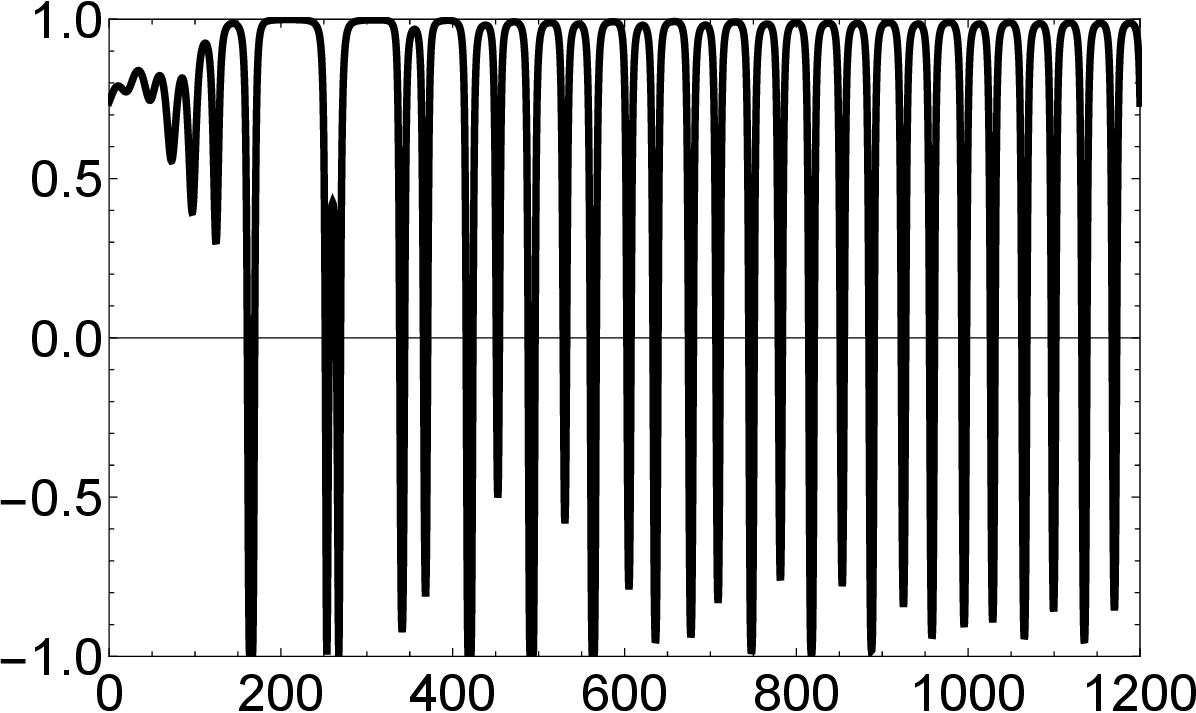}\\
     \quad a)&\quad b)
  \end{tabular}
  \caption{\label{fig:2} Evolution of angles between oscillators as the system (\ref{chimera}) evolves towards stable chimera state (for the same parameter values as in Figure \ref{fig:1}: (a) cosine of angle between a random oscillator from desynchronized sub-population and the synchronized sub-population and (b) cosine between two random oscillators from desynchronized sub-population.}
\end{figure*}

In whole, emergence of chimera in (\ref{chimera}) is unexpected and even counter-intuitive. Nevertheless, rigorous mathematical analysis based on the WS reduction demonstrates that such an equilibrium exists and it is stable (although the question of its stability within the full state space is quite subtle \cite{EM}). The dynamics in this equilibrium state look trivial at the first glance. Desynchronized sub-population is organized into TW that performs simple rotations, preserving its shape and the coherence. Hence, it looks like a stationary (up to rotations) equilibrium.
The angle between two sub-populations (i.e. between their densities) is constant.

In addition to TW as a whole (macroscopic dynamics), one can also observe the microscopic dynamics of individual oscillators. In Figure \ref{fig:2} we depict cosines between (randomly chosen) individual oscillators. Unexpectedly, it turns out that the angle between a random oscillator from desynchronized sub-population and the synchronized sub-population is not constant. Instead, it oscillates as shown in Figure \ref{fig:2}a). Figure \ref{fig:2}b) shows that an angle between two random oscillators from desynchronized sub-population also exhibits oscillations.

These findings seem surprising and indicate that simplicity of collective motions is deceptive.

We conclude that complicated motions of individual oscillators coalesce into simple rotations of their density. One might ask how is this possible. This question will be investigated in sections \ref{sec:3} and \ref{sec:4}. First, we make some preparations, by rewriting the system (\ref{chimera}) in new variables $z_j^l = e^{i \varphi_j^l}, \, l = A,B$
\begin{equation}
\label{Riccati_chimera}
\dot z_j^l = i (z_j^l)^2 f_l + i \omega z_j^l + i \bar f_l, \quad j=\overline{1,N}, \; l=A,B,
\end{equation}
where $f_A$ and $f_B$ are complex-valued coupling functions defined as
\begin{equation*}
\begin{split}
f_A &= \frac{i \mu}{2N} e^{i \beta} \sum \limits_{j=1}^{N} \bar z_j^A + \frac{i \nu}{2N} e^{i \beta} \sum \limits_{j=1}^{N} \bar z_j^B,\\
f_B &= \frac{i \mu}{2N} e^{i \beta} \sum \limits_{j=1}^{N} \bar z_j^B + \frac{i \nu}{2N} e^{i \beta} \sum \limits_{j=1}^{N} \bar z_j^A.
\end{split}
\end{equation*}
Notations $\bar w$ in the above equations stand for the conjugate of a complex number $w$.

Introduce centroids (centers of mass) of sub-populations $A$ and $B$:
$$
\zeta_A = \frac{1}{N} \sum \limits_{j=1}^N z_j^A, \quad \zeta_B = \frac{1}{N} \sum \limits_{j=1}^N z_j^B.
$$
Then the above expressions for coupling functions are simplified to
\begin{equation}
\label{coupling_chimera}
f_A = \frac{i}{2} e^{i \beta} (\mu \bar \zeta_A + \nu \bar \zeta_B), \quad f_B = \frac{i}{2} e^{i \beta} (\mu \bar \zeta_B + \nu \bar \zeta_A).
\end{equation}

Substituting (\ref{coupling_chimera}) into (\ref{Riccati_chimera}) yields the solvable chimera model (\ref{chimera}). Complex variables $z_j^A, \, z_j^B$ are more convenient for our further exposition, since such notations emphasize that we deal with the dynamical system on the circle (or on two-dimensional torus ${\mathbb T}^2 = S^1 \times S^1$, with one circle for each sub-population).

\subsection{TW in the conf-contr model}
\label{sec:2_2}
The second model describes two sub-populations with mixed (positive and negative) pairwise interactions. Oscillators from the first sub-population ("conformists") are attracted to all the others, while those belonging to the second sub-population ("contrarians") are repulsed by all the others. The governing equations for this model, named {\it the conformists-contrarians model}, read as \cite{HS2}
\begin{equation}
\label{conf-contr}
\left\{
\begin{split}
\textstyle{\dot \varphi_j^C=}& \textstyle{\omega + \frac{K_C}{M+N} \left( \sum \limits_{i=1}^M \sin (\varphi_j - \varphi^C_i) + \sum \limits_{i=1}^N \sin (\varphi_j - \varphi_i^D) \right)},\\
&\textstyle{j=1,\dots,M;} \\
\textstyle{\dot \varphi_j^D=}& \textstyle{\omega + \frac{K_D}{M+N} \left( \sum \limits_{i=1}^M \sin (\varphi_j - \varphi^C_i) + \sum \limits_{i=1}^N \sin (\varphi_j - \varphi_i^D) \right)},\\
&\textstyle{j=1,\dots,N,}
\end{split}\right.
\end{equation}
where $K_C > 0$ and $K_D < 0$. Hence, there are $M+N$ oscillators in total, out of which there are $M$ "conformists" and $N$ "contrarians", denoted by subscripts $C$ and $D$, respectively. Denote by $p=M/(M+N)$ the portion of conformists in the whole population, and by $q = N/(M+N) = 1-p$ the portion of contrarians. Underline that all oscillators are assumed to have identical intrinsic frequencies.
\begin{figure}[t]
\centering
  \begin{tabular}{@{}c@{}}
    \includegraphics[width=.4\textwidth]{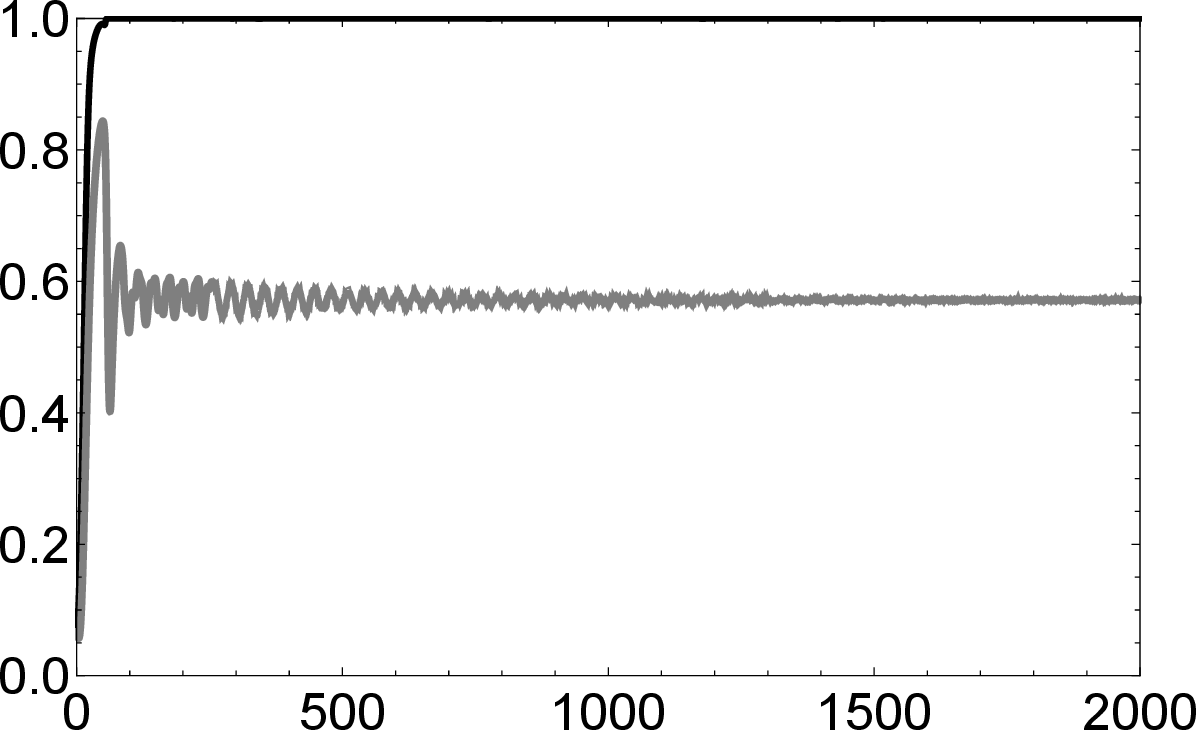}
  \end{tabular}
  \caption{\label{fig:3} Evolution of real order parameters $r_C$ and $r_D$ as the system (\ref{conf-contr}) evolves towards the traveling wave state. Obtained by solving (\ref{conf-contr}) with $N = 500$ oscillators and parameter values $p=0.5$, $K_C = 0.7$, $K_D =-0.3$.}
\end{figure}

The model (\ref{conf-contr}) has been analytically studied using the WS reduction \cite{HS2}. It has been proven that (\ref{conf-contr}) admits four different equilibria depending on two parameters $K_C/K_D$ and $p$. The most interesting is TW, the stable equilibrium state in which conformists are perfectly synchronized ($r_C = 1$), while contrarians are only partially synchronized with constant real order parameter $0< r_D < 1$. Both sub-populations rotate along the circle, preserving the constant angle between them. In other words, the angle $\delta$ between synchronized conformists and the peak of TW consisting of contrarians is constant. In addition, this is the only equilibrium in (\ref{conf-contr}) which is not antipodal: equilibrium value of the angle $\delta$ is strictly less than $\pi$. Figure \ref{fig:3} demonstrates the evolution of real order parameters $r_C$ and $r_D$ as the system evolves towards the TW state.

\begin{figure*}[t]
\centering
  \begin{tabular}{@{}cc@{}}
    \includegraphics[width=.4\textwidth]{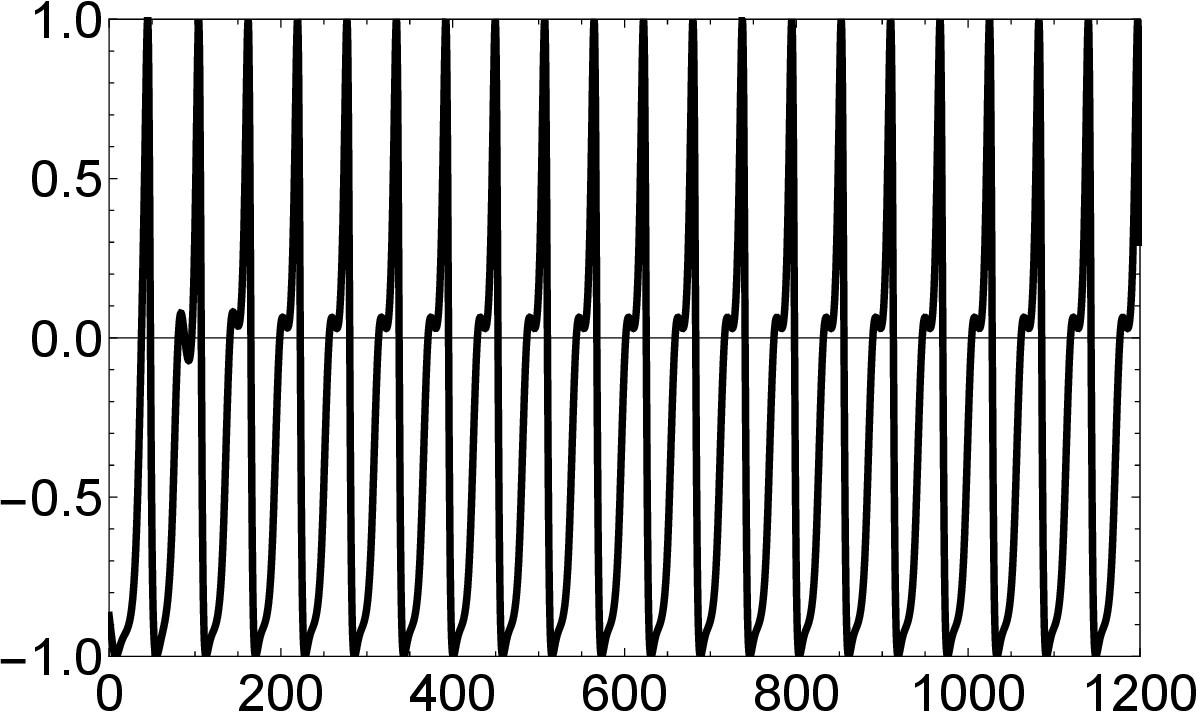} &
    \includegraphics[width=.4\textwidth]{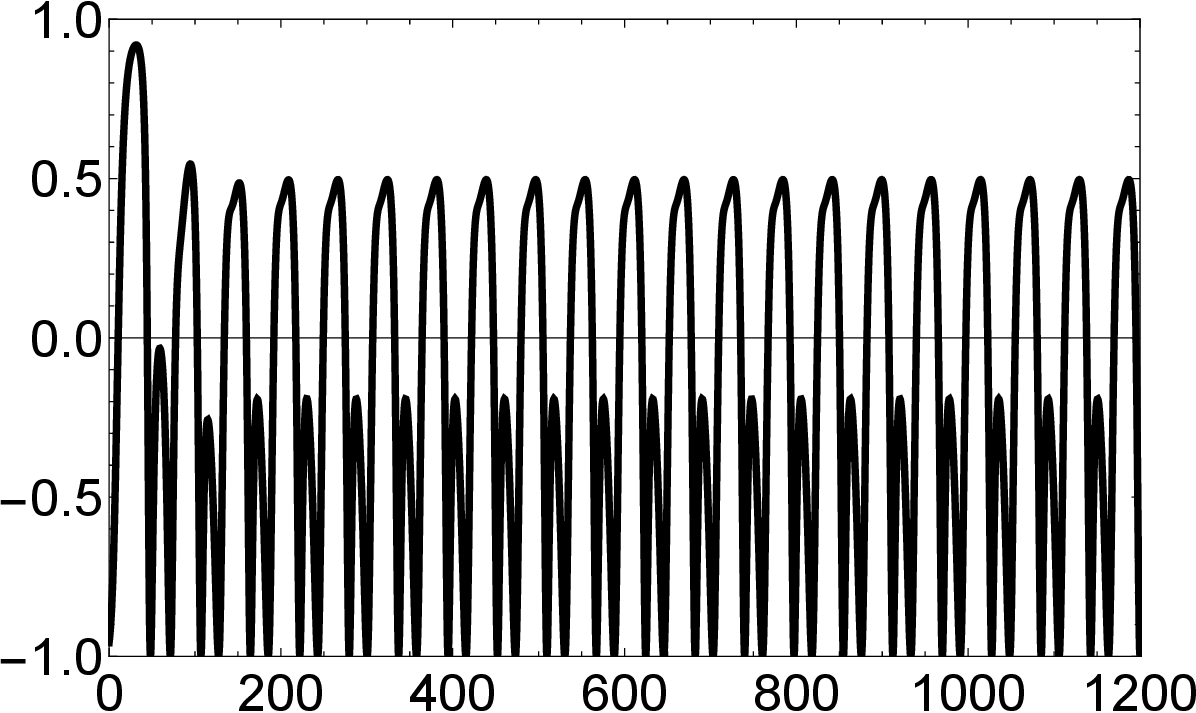}\\
     \quad a)&\quad b)
  \end{tabular}
  \caption{\label{fig:4} Evolution of angles between oscillators as the system (\ref{conf-contr}) evolves towards the traveling wave state (for the same parameter values as in Figure \ref{fig:3}: (a) cosine of the angle between a random contrarian and conformist and (b) cosine of the angle between two random contrarians.}
\end{figure*}

In order to get the first insight into this equilibrium state, we simulate (\ref{conf-contr}) and observe (cosines of) angles between random oscillators. The oscillatory dynamics shown in Figure \ref{fig:4} indicates once again that microscopic dynamics (individual contrarians) are more complicated than macroscopic ones (density of contrarians).

For further analysis it is convenient to pass to complex variables $z_j^l = e^{i \varphi_j^l}, \, l=C,D$ and rewrite the system (\ref{conf-contr}) as
\begin{equation}
\label{Riccati_conf-contr}
\dot z_j^l = i (z_j^l)^2 f_l + i \omega z_j^l + i \bar f_l, \quad j=\overline{1,N}, \; l=C,D,
\end{equation}
where $f_C$ and $f_D$ are complex-valued coupling functions given by:
\begin{equation}
\label{coupling_conf-contr}
\begin{split}
f_C &= \frac{i K_C}{2}(p \bar \zeta_C + (1-p) \bar \zeta_D),\\
f_D &= \frac{i K_D}{2} (p \bar \zeta_C + (1-p) \bar \zeta_D),
\end{split}
\end{equation}
with $K_C > 0, K_D<0.$

\section{Dynamics on the M\" obius group and its orbits}
\label{sec:3}
Analytic investigation of TW's in (\ref{chimera}) and (\ref{conf-contr}) relies on reduction of these systems to low-dimensional dynamics. Consider the group of M\" obius transformations (conformal mappings) of the complex plane. Let ${\mathbb G}$ be the subgroup of those M\" obius transformations that leave the unit disc invariant. This subgroup consists of isometries (in hyperbolic metric) of the unit disc. It is isomorphic to the matrix group $PSU(1,1) = SU(1,1) / \pm I$.
We will refer to the (sub)group ${\mathbb G}$ as {\it M\" obius group}. Transformations from this group can be written in the following form
\begin{equation}
\label{Mobius}
g(z) = \frac{e^{i \psi} z + \alpha}{1 + \bar \alpha e^{i \psi} z}, \mbox{  where } z \in {\mathbb C}, \, |z| \leq 1.
\end{equation}
Parameters of this transformation are the angle $\psi \in [0,2 \pi]$ and the complex number $\alpha \in {\mathbb C}, \, |\alpha| < 1$. Real dimension of the group manifold of ${\mathbb G}$ equals three. The WS result reduces the collective dynamics in large ensembles to the system of ODE's for global variables $\alpha$ and $\psi$. In other words, globally coupled ensemble of identical oscillators evolves by actions of the M\" obius group \cite{MMS}. By slightly adapting this result we assert that each of four sub-populations in models (\ref{chimera}) and (\ref{conf-contr}) evolve by actions of the M\" obius group. We substantiate this in the following two propositions.

\begin{proposition}
\label{prop:1}
Consider the chimera model (\ref{Riccati_chimera}).
There exist two one-parametric families $g_t^A, \, g_t^B \in {\mathbb G}$, such that
$$
z_j^A(t) = g_t^A(z_j^A(0)), \mbox{  and  } z_j^B = g_t^B(z_j^B(0)) \quad \forall j=\overline{1,N}.
$$
Furthermore, parameters $\alpha_A, \psi_A, \alpha_B, \psi_B$ of the families $g_t^A$ and $g_t^B$ satisfy the following systems of ODE's
\begin{equation}
\label{WS^2}
\left \{
\begin{array}{ll}
\dot \alpha_l = i (f_l \alpha_l^2 + \omega \alpha_l + \bar f_l); \\
\dot \psi_l = f_l \alpha_l + \omega + \bar f_l \bar \alpha_l, \\
\end{array}
\right.
\end{equation}
where the coupling functions $f_A$ and $f_B$ are given by (\ref{coupling_chimera}).
\end{proposition}
\begin{proposition}
\label{prop:2}
Consider the conformists-contrarians model (\ref{Riccati_conf-contr}).
There exist two one-parametric families $g^C_t, g^D_t \in {\mathbb G}$, such that
$$
z_j^C(t) = g_t^C(z_j^C(0)), \mbox{  and  } z_j^D = g_t^D(z_j^D(0)) \quad \forall j=\overline{1,N}.
$$
Furthermore, parameters $\alpha_C, \psi_C, \alpha_D, \psi_D$ of the families $g_t^C$ and $g_t^D$ satisfy the systems of ODE's (\ref{WS^2}), where the coupling functions $f_C$ and $f_D$ are given by (\ref{coupling_conf-contr}).
\end{proposition}

Propositions \ref{prop:1} and \ref{prop:2} state that each of systems (\ref{chimera}) and (\ref{conf-contr}) generate a trajectory in the Lie group ${\mathbb G} \times {\mathbb G}$. The equations in (\ref{WS^2}) are mutually coupled through complex-valued functions $f_A, f_B$ in the first model and $f_C, f_D$ in the second. Therefore, the system evolves on orbits of the group ${\mathbb G} \times {\mathbb G}$. Since the corresponding group manifold is six-dimensional, the evolution in each model takes place on a six-dimensional invariant submanifold (three-dimensional invariant submanifold for each sub-population). This six-dimensional submanifold is determined by the initial state of the system (i.e. by initial distribution of oscillators).

\subsection{Reduced dynamics on the Poisson manifold}
\label{sec:3_1}
We will be interested in the special case when dynamics are further reduced to four-dimensional invariant submanifolds (that is - to two-dimensional submanifold for each sub-population). This occurs if the initial distribution of oscillators is uniform, as substantiated in the following

\begin{proposition}
\label{prop:3}
[\onlinecite{MMS}] Consider models (\ref{chimera}) and (\ref{conf-contr}) in thermodynamic limit $N \to \infty$.

Suppose that initial distributions of oscillators belonging to all sub-populations in models (\ref{chimera}) and (\ref{conf-contr}) are uniform on the unit circle $S^1$. (In other words, we assume that initial distribution of all sub-populations are given by the density function $\rho_l(0,\varphi) = 1 / 2 \pi$ for $l=A,B,C,D$.)

Then, distributions of oscillators in sub-populations $l=A,B,C,D$ at each moment $t$ are given by the following density functions
\begin{equation}
\label{Poisson}
\rho_l(t,\varphi) = \frac{1}{2 \pi} \frac{1-r_l^2(t)}{1 - 2 r_l(t) \cos(\varphi - \Phi_l(t)) + r_l^2(t)}.
\end{equation}
Here, $\alpha_l = r_l(t) e^{i \Phi_l(t)}, \, 0 \leq r_l(t) < 1$ and $0 \leq\varphi \leq 2 \pi$.
\end{proposition}

\begin{remark}
\label{rem:1}
Functions of the form (\ref{Poisson}) are classical objects in complex analysis and potential theory, well known as {\it Poisson kernels}. We treat (\ref{Poisson}) as densities of probability distributions on the unit circle, parametrized by a point $\alpha_l = r_l e^{\Phi_l}$ in the unit disc $\mathbb{B}^2$. \footnote{In directional statistics probability distributions with densities (\ref{Poisson}) are named {\it wrapped Cauchy distributions}\cite{McCullagh}.} Therefore, the dimension of the invariant submanifold of probability measures on $S^1$ with densities (\ref{Poisson}) equals two. We refer to this submanifold as {\it Poisson manifold}.
\end{remark}

\begin{remark}
\label{rem:2}
Uniform measure and delta distributions are two extreme cases of Poisson kernels, obtained for $r_l = 0$ and the limit case $r_l \to 1$, respectively.
\end{remark}

\begin{remark}
\label{rem:3}
Complex number $\alpha_l$ is the mean value (complex order parameter) for the density (\ref{Poisson}) and $r_l = |\alpha_l|$ is the real order parameter for (\ref{Poisson}).
\end{remark}

\section{Phase holonomy in two-populations Kuramoto models: simulations}
\label{sec:4}
Group ${\mathbb G}$ operates on the unit circle $S^1$, on unit disc $\mathbb{B}^2$ and on space ${\cal P}(S^1)$ of probability measures on the circle. Explicitly
$$
g \cdot z = g(z), \mbox{ if }z \in S^1 \cup \mathbb{B}^2;
$$
$$
g \cdot \mu (A) = g_* \mu(A) = \mu(g^{-1}(A)),
$$
if $\mu \in {\cal P}(S^1)$ and $A \subseteq S^1$ is a Borel set. Hence, $g_* \mu$ denotes a measure obtained by the action of $g \in {\mathbb G}$ on the measure $\mu$.

The state of sub-population $l=A,B,C,D$ at a moment $t$ is represented by a probability measure $\mu_l(t) \in {\cal P}(S^1)$. Let $\mu_l(0)$ be the initial state. From propositions \ref{prop:1} and \ref{prop:2} it follows that the state at each moment $t$ is a certain M\" obius transformation of the initial state:
$$
\mu_l(t) = g^l_{t \; *} \mu_l(0).
$$
Moreover, let $\mu_l(t_1)$ and $\mu_l(t_2)$ be states of the sub-population $l$ at moments $t_1$ and $t_2$, respectively. Then there exists a M\" obius transformation that maps $\mu_l(t_1)$ into $\mu_l(t_2)$. Denote this transformation by $g^l_{[t_1,t_2]}$. Using this notation we can write:
$$
\mu_l(t_2) = g^l_{[t_1,t_2] \; *} \mu_l(t_1).
$$
Notice that $g_{[t,t]}$ is the identity transformation for all $t$.
\begin{figure}[t]
\centering
  \begin{tabular}{@{}c@{}}
    \includegraphics[width=.4\textwidth]{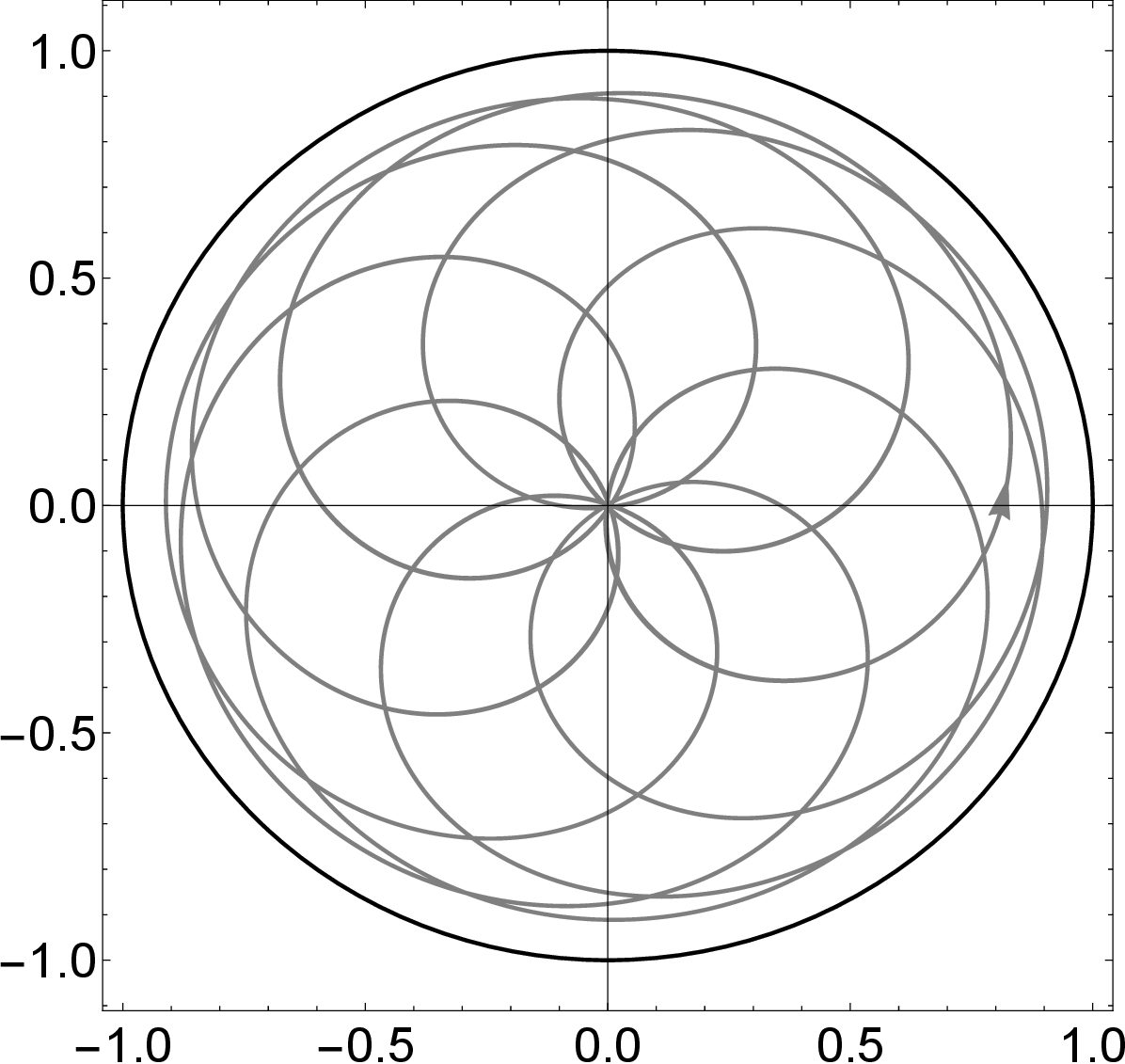}
  \end{tabular}
  \caption{\label{fig:5} Path $\alpha(2000;t)$ in the unit disc for the stable chimera state in (\ref{chimera}). Obtained for the same parameter values as in Figure \ref{fig:1}.}
\end{figure}

In these refined notations the transformation $g_t^l$ is denoted as $g_{[0,t]}^l$.
It is obvious from (\ref{Mobius}) that parameters $\alpha_l, \; l=A,B,C,D$ are images of zero under maps $g^l_{[0,t]}$, that is:
$$
\alpha_l(t) = g^l_{[0,t]}(0), \quad l=A,B,C,D.
$$
Finally, denote by $\alpha_l(t_1;t)$ images of zero under the maps $g^l_{[t_1,t]}$:
$$
\alpha_l(t_1;t) = g^l_{[t_1,t]}(0), \quad l=A,B,C,D.
$$

\begin{remark}
\label{rem:4}
It is obvious from (\ref{Mobius}) that $g^l_{[t_1,t]}$ is a simple rotation if and only if $\alpha_l(t_1;t) = 0$.
\end{remark}

\subsection{Phase holonomy in the solvable chimera model: simulations}
\label{sec:4_1}
Here, we consider the stable chimera state in the model (\ref{chimera}). Fully synchronized sub-population is denoted by $A$, and desynchronized sub-population by $B$. Assuming that initial distributions are uniform, Proposition \ref{prop:3} implies that densities of both sub-populations are given by (\ref{Poisson}), with $r_A(t) \to 1$ and $0 < r_B(t) \to const < 1$ as $t \to \infty$.

At the first glance, the stable chimera state is stationary up to rotations, as the complex order parameter $\alpha_B$ evolves along the circle inside the unit disc ${\mathbb B}^2$. However, simulation results shown in Figure \ref{fig:2} indicate that these collective motions are deceptively simple.

In order to get a better insight, fix sufficiently large $T_1$, so that $r_B(t) = const$ for $t > T_1$. At each moment $t \geq T_1$, based on positions of three oscillators, calculate the transformation $g^B_{[T_1,t]}$ that acts on $\mu_B(T_1)$. We observe the image of zero under the transformation $g^B_{[T_1,t]}$. The path $\alpha_B(T_1;t) = g^B_{[T_1,t]}(0)$ is depicted in Figure \ref{fig:5}.

If transformations acting on $\mu_B(T_1)$ would be simple rotations, then $\alpha(T_1,t)$ would stay at zero at all times $t$. Figure \ref{fig:5} demonstrates that this is not the case. However, one can notice that $\alpha(T_1;t)$ repeatedly returns to zero. We conclude that individual oscillators evolve by transformations that are not simple rotations, but there exists a sequence of moments $T_2,T_3,...$, at which transformations $g^B_{[T_1,T_2]},g^B_{[T_1,T_3]},...$ are simple rotations.

\subsection{Phase holonomy in the conf-contr model: simulations}
\label{sec:4_2}
We further examine the TW state in the conf-contr model (\ref{conf-contr}). In this equilibrium state conformists are fully synchronized, while contrarians remain only partially synchronized. Moreover, both sub-populations travel around the circle, preserving the constant angle $\delta < \pi$ between them.

It is assumed that initial distributions are uniform. Then, due to Proposition \ref{prop:3}, densities at each moment $t$ are of the form (\ref{Poisson}). In equilibrium, real order parameters are constant: $r_C = 1$ and $0<r_D = const < 1$.

We are interested in dynamics on the M\" obius group ${\mathbb G}$ which corresponds to the sub-population of contrarians. Fix sufficiently large $T_1$, so that $r_D(t) = const$ for $t>T_1$. Denote by $g^D_{[T_1,t]}$ the transformation that acts on measure $\mu_D(T_1)$ on time interval $[T_1,t]$.

Figure \ref{fig:6} shows the path of $\alpha_D(T_1;t) = g^D_{[T_1,t]}(0)$ in the unit disc for $t_1=2000$. Observe that $\alpha_D(T_1,t)$ also repeatedly returns to zero. This means that contrarian sub-population in the TW state does not perform simple rotations, but there exists a sequence of moments $T_1, T_2,...$ for which $g^D_{[T_1,T_2]}, g^D_{T_1,T_3]},...$ are simple rotations.

\subsection{Holonomy associated with TW's of Kuramoto oscillators: discussion}
\label{sec:4_3}
\begin{figure}[t]
\centering
  \begin{tabular}{@{}c@{}}
    \includegraphics[width=.4\textwidth]{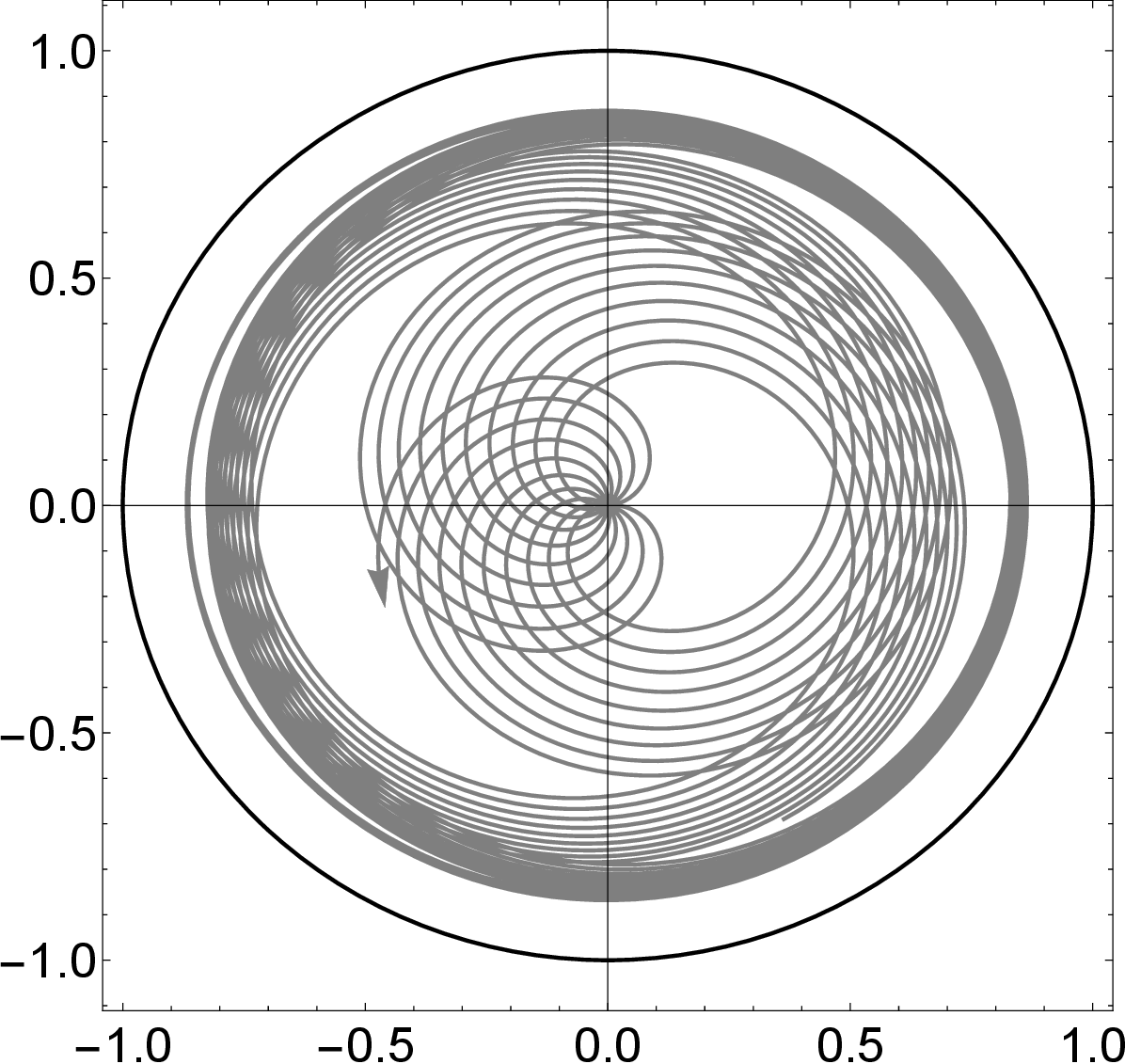}
  \end{tabular}
  \caption{\label{fig:6} Path $\alpha(2000;t)$ in the unit disc for the traveling wave state in (\ref{conf-contr}). Obtained for the same parameter values as in Figure \ref{fig:3}.}
\end{figure}

Figures \ref{fig:5} and \ref{fig:6} confirm that cyclic evolution of TW's in (\ref{chimera}) and (\ref{conf-contr}) is associated with the phase shift. Mathematically, this is explained as a holonomy on fiber bundle of the group $SU(1,1)$. The configuration space is ${\mathbb G}$, and the base space is the parameter space for the Poisson manifold, which is identified with unit disc ${\mathbb B}^2$. This corresponds to the decomposition $G/U(1) = {\mathbb B}^2$, with fiber bundle consisting of elements of the group $U(1)$, that is - of circles. Motions on orbits of the M\" obius group are restricted to circles inside ${\mathbb B}^2$. However, horizontal lifts of these circles are not closed loops in ${\mathbb G}$.
In fact, circles in ${\mathbb B}^2$ traced by $\alpha_l, \, l=B,D$ are projections of trajectories restricted on tori in the group ${\mathbb G}$.

The difference between motifs depicted in figures 5 and 6 unveils that parallel translations for a connection on the fiber bundle ${\mathbb G}$ act differently on fibers for the chimera state and TW.

\section{Phase holonomy in two-populations Kuramoto models: analysis}
\label{sec:5}
Now, we pass to the analytic study of the phase holonomy in two models. Starting from a certain moment $T_1$ the real order parameters of desynchronized sub-populations are constant: $r_B(t_2) = r_B(t_1)$ and $r_D(t_2) = r_D(t_1)$ for all $t_1, t_2 > T_1$. Therefore, densities perform cyclic evolutions.

Fix $\tau>T_1$ and denote by $T$ the time interval of one cyclic evolution starting from $\tau$ (note that the evolutions are not periodic, hence, $T$ depends on $\tau$). Using notations from the previous Section, we can state it as follows
$$
\mu(\tau + T) = g_{[\tau,\tau+T] \; *}^l  \mu(\tau), \quad l=B,D.
$$
In other words, $\rho_l(\tau+T,\varphi) = \rho_l(\tau,\varphi)$, so $\Phi_l(\tau+T) = \Phi_l(\tau)$ for $l=B,D$.

Our goal in this Section is to evaluate the phase shift $\psi_l(\tau+T) - \psi_l(\tau)$ during one cycle.

\subsection{Phase holonomy in the solvable chimera model: analysis}
\label{sec:5_1}
We first analyze the system (\ref{chimera}) at time interval $t \in [\tau,\tau+T]$.

For the whole system (i.e. for both sub-populations) the dynamics is restricted to the product ${\mathbb G} \times {\mathbb G}$ and given by the system of ODE's (\ref{WS^2}). For $t > T_1$ these dynamics are restricted to an invariant 3-torus ${\mathbb T}^3 \subset {\mathbb G} \times {\mathbb G}$. This torus is parametrized by angles $\Phi_A, \Phi_B$ and $\psi_B$. Since $r_A(t) = 1$ and $r_B(t) = const$ for $t>\tau$, we have that $\alpha_A(t) = e^{i \Phi_A(t)}$ and $\alpha_B(t) = r_B e^{i \Phi_B(t)}$. Substituting into (\ref{WS^2}) and discarding the equation for $\psi_A$, we obtain the dynamical system on ${\mathbb T}^3$
\begin{equation}
\label{3-torus_chimera}
\left\{
\begin{array}{lll}
\dot \Phi_A = f_A e^{i \Phi_A} + \omega + \bar f_A e^{-i \Phi_A};\\
\dot \Phi_B = r_B f_B e^{i \Phi_B} + \omega + \frac{1}{r_B} \bar f_B e^{-i \Phi_B};\\
\dot \psi_B = r_B f_B e^{i \Phi_B} + \omega + r_B \bar f_B e^{-i \Phi_B}.
\end{array}
\right.
\end{equation}
We will focus on the dynamics on an invariant 2-torus ${\mathbb T}^2 \subset {\mathbb G}$ which is described by angles $\Phi_B$ and $\psi_B$ corresponding to the desynchronized sub-population.

Substitute expressions (\ref{coupling_chimera}) for $f_A$ and $f_B$ into (\ref{3-torus_chimera}). Coupling functions $f_A$ and $f_B$ depend on centroids $\zeta_A$ and $\zeta_B$. However, taking into account that we consider dynamics on the Poisson manifold, centroids can be replaced by $\alpha_A$ and $\alpha_B$, respectively (see Remark \ref{rem:3} in Section \ref{sec:3}). Imposing that the right hand side of ODE's for $\Phi_A$ and $\Phi_B$ must be real, we find that
\begin{equation}
\label{expr_chimera}
\nu \cos (\Phi_A - \Phi_B - \beta) + r_B \mu \cos \beta = 0.
\end{equation}
Equality (\ref{expr_chimera}) is the necessary condition for the stable chimera to exist \cite{AMSW}. Then the system on ${\mathbb T}^2$ is rewritten as
\begin{equation}
\label{2-torus_chimera}
\left\{
\begin{split}
\dot \Phi_B =& - \frac{\mu}{2}(r_B^2 + 1) \sin \beta + \omega+\\
&+\frac{\nu}{2}(r_B + \frac{1}{r_B}) \sin(\Phi_A - \Phi_B - \beta);\\
\dot \psi_B =& - \mu r_B^2 \sin \beta + \omega + \nu r_B \sin(\Phi_A - \Phi_B - \beta).
\end{split}
\right.
\end{equation}
By subtracting the first equation in (\ref{2-torus_chimera}) from the second, we obtain ODE for the phase difference
$$
\frac{d}{dt}(\psi_B - \Phi_B) = (r_B - \frac{1}{r_B}) (\nu \sin(\Phi_A - \Phi_B - \beta) - r_B \mu \sin \beta).
$$
Using (\ref{expr_chimera}) and trigonometric identities, the above ODE can be rearranged to get
$$
\frac{d}{dt}(\psi_B - \Phi_B) = \left( r_B - \frac{1}{r_B} \right) \frac{\nu}{\cos \beta} \sin(\Phi_A - \Phi_B).
$$
Finally, integration over the time interval $[\tau,\tau+T]$ yields
\begin{equation}
\label{geom_phase_chimera}
\begin{split}
&(\psi_B(\tau+T) - \Phi_B(\tau+T)) - (\psi_B(\tau) - \Phi_B(\tau)) =\\
&=\left( r_B - \frac{1}{r_B} \right) \frac{\nu}{\cos \beta} \int \limits_\tau^{\tau+T} \sin(\Phi_A - \Phi_B) dt.
\end{split}
\end{equation}
It follows from (\ref{expr_chimera}) that the difference $\Phi_A - \Phi_B$ is constant. Taking this into account, along with the fact that $\Phi_B(\tau+T) = \Phi_B(\tau)$, (\ref{geom_phase_chimera}) is rewritten as
\begin{equation}
\label{geom_phase_chimera_2}
\psi_B(\tau+T) - \psi_B(\tau) = \left( r_B - \frac{1}{r_B} \right) \frac{\nu T}{\cos \beta} \sin(\Phi_A - \Phi_B).
\end{equation}
Equality (\ref{geom_phase_chimera_2}) gives the exact expression for the phase shift in the stable chimera state during one cycle which starts at the moment $\tau > T$ (recall that $T$ depends on $\tau$).

\subsection{Phase holonomy in the conf-contr model: analysis}
\label{sec:5_2}
We now consider the conf-contr model (\ref{conf-contr}). Just as in the previous case, the WS dynamics for $t > T_1$ are restricted on invariant 3-torus ${\mathbb T}^3 \subset {\mathbb G} \times {\mathbb G}$, parametrized by angles $\Phi_C, \Phi_D$ and $\psi_D$, where the subscript $D$ corresponds to the contrarian sub-population. The system of ODE's is the same as (\ref{3-torus_chimera}), with the single difference that subscripts $A$ and $B$ are replaced by $C$ and $D$, respectively.

Further, substitute the expressions (\ref{coupling_conf-contr}) for $f_C$ and $f_D$ (replacing $\zeta_C$ and $\zeta_D$ by $\alpha_C$ and $\alpha_D$). Equating imaginary parts of the right hand sides of ODE's to zero we get
\begin{equation}
\label{expr-conf-contr}
\cos (\Phi_D - \Phi_C) = r_D \left( 1 - \frac{1}{p}\right)
\end{equation}
which is precisely the necessary condition for the TW state \cite{HS2}.

 Then the system for variables $\Phi_D$ and $\psi_D$ is rewritten as
\begin{equation}
\label{2-torus_conf-contr}
\left\{
\begin{array}{rcl}
\dot \Phi_D &=& \left(r_D + \frac{1}{r_D} \right) \frac{p K_D}{2} \sin (\Phi_D - \Phi_C) + \omega \\
\\
\dot \psi_D &=& p K_D r_D \sin(\Phi_C - \Phi_D) + \omega
\end{array}
\right.
\end{equation}
and the difference between two phases satisfies
$$
\frac{d}{dt}(\psi_D - \Phi_D) = \frac{p K_D}{2} \left(r_D + \frac{1}{r_D} \right) \sin(\Phi_C - \Phi_D).
$$
Finally, integration of the last ODE over the time cycle $[\tau,\tau+T]$ yields
\begin{equation}
\label{geom_phase_C-C}
\begin{split}
&(\psi_D(\tau+T) - \Phi_D(\tau+T)) - (\psi_D(\tau) - \Phi_D(\tau)) = \\
&=\frac{p K_D}{2} \left(r_D + \frac{1}{r_D} \right) \int \limits_{\tau}^{\tau+T}\sin(\Phi_C - \Phi_D) dt.
\end{split}
\end{equation}
Just as in the previous subsection, taking into account that $\Phi_D - \Phi_C$ is constant and that $\Phi_D(\tau+T) = \Phi_D(\tau)$, we find that
\begin{equation}
\label{geom_phase_C-C_2}
\psi_D(\tau+T) - \psi_D(\tau) = \frac{p T K_D}{2} \left(r_D + \frac{1}{r_D} \right) \sin(\Phi_C - \Phi_D).
\end{equation}
This is the expression for the phase shift during one cycle of TW in the conf-contr model (\ref{conf-contr}) starting from the moment $\tau$.

\section{Phase holonomy in the single populaton model}
\label{sec:6}
In sections \ref{sec:4} and \ref{sec:5} we have exposed that unexpected stable equilibria in two-populations Kuramoto networks are associated with phase holonomy. However, Kuramoto networks with two sub-populations are not minimal models that exhibit such an effect. In fact, phase holonomy arises even in the simplest setup, with a single population of identical globally coupled oscillators of the form
\begin{equation}
\label{global-coupling}
\dot \varphi_j = \omega + \frac{K}{N} \sum \limits_{i=1}^N \sin(\varphi_i-\varphi_j - \beta), \quad j=1,\dots,N.
\end{equation}
This system generates a trajectory on the M\" obius group ${\mathbb G}$ which depends on the initial distribution of oscillators' phases \cite{MMS}.
Uniform initial distribution of oscillators yields evolution of
densities on the two-dimensional invariant Poisson manifold. The subtlety of this exceptional case has been briefly pointed out previously \cite{MMS}. When dynamics are restricted on the Poisson manifold, ODE's for global variables $\alpha$ and $\psi$ decouple and dynamics of $\psi$ becomes seemingly irrelevant.
\begin{figure}[t]
\centering
  \begin{tabular}{@{}c@{}}
    \includegraphics[width=.4\textwidth]{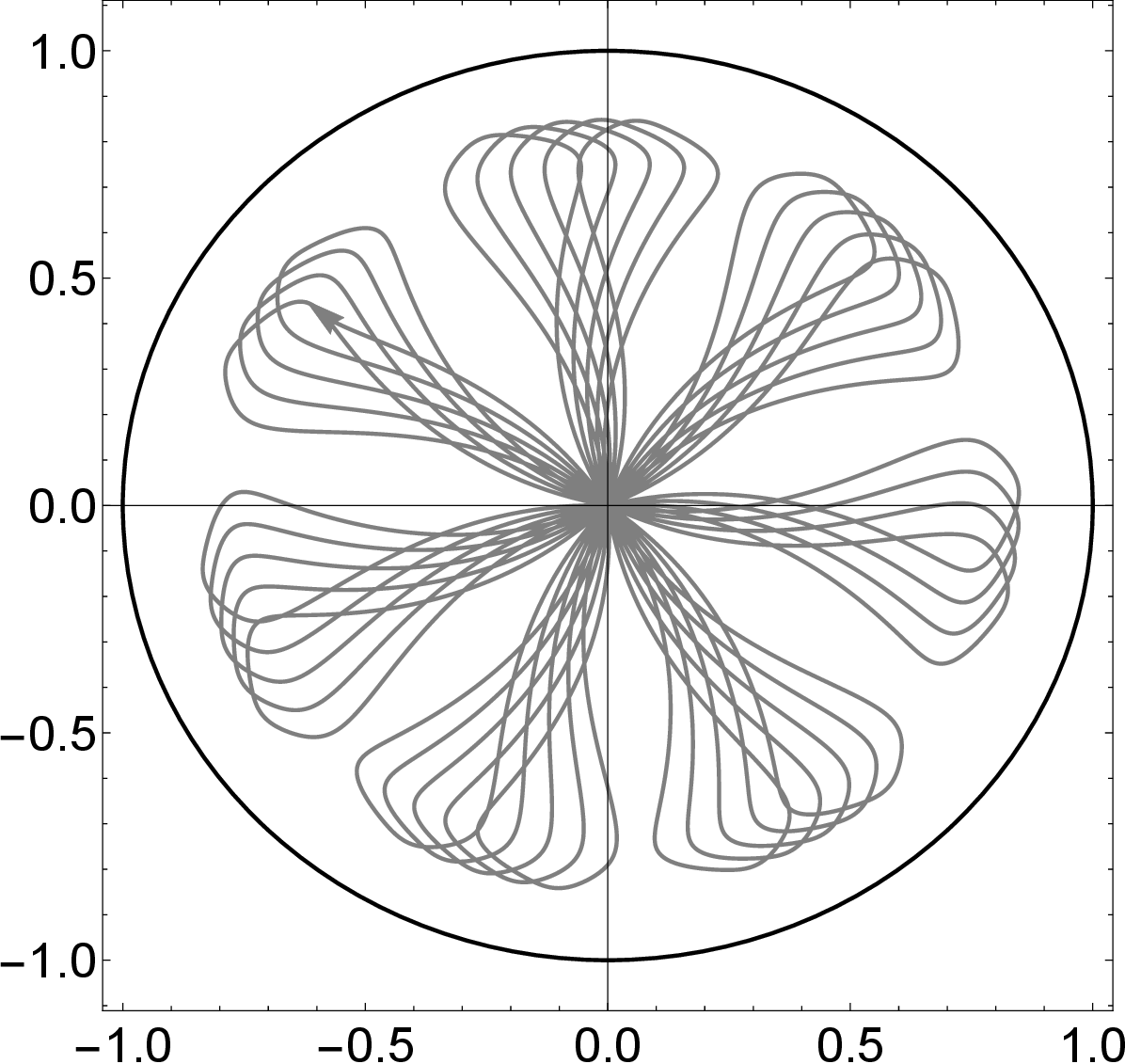}
  \end{tabular}
  \caption{\label{fig:7} Path $\alpha(500;t)$ in the unit disc for the model (\ref{global-coupling}). Obtained by solving (\ref{global-coupling}) for $N = 500$ oscillators and parameter values $K = 1$ and $\beta = \frac{\pi}{2}$.}
\end{figure}

We are interested in cyclic evolution on the Poisson manifold.
It has been shown that systems of the form (\ref{global-coupling}) induce gradient flows in the unit disc w. r. to the hyperbolic metric, where the potential has a unique critical point \cite{CEM}. Closed trajectories are ruled out for all values of the phase shift $\beta$ not equal to $\pi/2$. The particular model (\ref{global-coupling}) with $\beta=\pi/2$ yields Hamiltonian dynamics in the unit disc.

Hence, in order to obtain cyclic evolution on the Poisson manifold, we choose a non-uniform initial density of the form (\ref{Poisson}) with $0<r<1$. Further, we solve (\ref{global-coupling}) with $\beta = \pi/2$ for chosen initial conditions.
This yields cyclic evolution of Poisson densities (\ref{Poisson}), with the constant real order parameter $r(t)$. Denote by $\alpha_{[t_1,t]}$ an image of the zero under the M\" obius transformation acting on the density at the time interval $[t_1,t]$. Path of $\alpha_{[500,t]}$ in the unit disc is plotted in Figure \ref{fig:7}. This Figure confirms that M\" obius transformations generated by (\ref{global-coupling}) are not simple rotations. Therefore, more complicated dynamics of individual oscillators take place under the profile of one-dimensional cyclic evolution of the density. This further means that the closed loop (circle inside the unit disc) traced by the complex order parameter $\alpha(t) = r(t) e^{i \Phi(t)}$ is the projection of an open trajectory in the group ${\mathbb G}$.

Calculations analogous to those presented for the model with two sub-populations in Section \ref{sec:5} yield an explicit expression for the phase shift during cyclic evolution over the time interval $[\tau,\tau+T]$:
$$
\psi(\tau + T) - \psi(\tau) = - \frac{KT}{2}(1+r^2).
$$

\section{Analogies with cyclic quantum evolution on submanifold of coherent states}
\label{sec:7}
At the early stage of mathematical quantum-mechanic theory, Erwin Schr\" odinger investigated quantum states whose evolution resembles oscillatory dynamics of classical harmonic oscillator \cite{Schrodinger}. Following Schr\" odinger, these quantum states are named {\it coherent states}. \footnote{In order to avoid confusion, underline that the notion of coherent states in quantum physics has nothing in common with the term "coherent state" that refers to the full synchronization of coupled oscillators.}

During XX century, general mathematical approach to coherent states based on group theory and irreducible representations, was developed by Gilmore \cite{Gilmore} and Perelomov \cite{Perelomov}. We refer to the book \cite{Gazeau} for an overview of coherent states in mathematical physics. Coherent states constitute a submanifold that is invariant for quantum evolution, given that the Hamiltonian is a linear combination of generators of a certain group.

Geometric phase has been frequently studied in the context of cyclic quantum evolution on coherent states, as such a setup allows for transparent expressions \cite{CSS,FA,LHC}.

In the previous sections we investigated cyclic evolution on the Poisson manifold. Here, we briefly point out the analogy with the evolution on manifold of coherent states.

\subsection{Poisson kernels as $SU(1,1)$ coherent states. TW's in Kuramoto models as cyclic evolution on submanifold of coherent states}
\label{sec:7_1}
Poisson kernels sometimes appear as coherent states in mathematical physics. For instance, in some quantum systems certain classes of states are represented by probability distributions on the circle. Such states are called {\it phase-like quantum states} \cite{Wunsche}. Phase-like quantum states represented by Poisson kernels are coherent states. Another mathematical framework is provided by representations of quantum states with analytic complex functions \cite{Vourdas}. In the case of hyperbolic geometry, the manifold of coherent states is precisely the Poisson manifold, seen as analytic functions on unit disc in the complex plane. These states play an important role in quantum optics \cite{dMSV}.

TW's in models (\ref{chimera}) and (\ref{conf-contr}) can be regarded as a cyclic evolution of probability densities, with the system returning to an initial state.

Furthermore, interpretation of the Poisson manifold as coherent states fits into Perelomov's general mathematical framework of coherent states \cite{Perelomov}. Poisson kernels are hyperbolic (sometimes also referred to as {\it pseudo-spin} or $SU(1,1)$-) coherent states \cite{CR}. These states correspond to the decomposition $SU(1,1) / U(1) = {\mathbb B}^2$, with $U(1)$ being the maximal compact subgroup of $SU(1,1)$. \footnote{On the other hand, the quantum spin corresponds to the decomposition $SU(2)/U(1) = {\mathbb S}^2$, where ${\mathbb S}^2$ is the two-sphere. Geometric phase (AA phase) for the spin coherent states arise for cyclic evolution (parallel transport along a closed loop) in ${\mathbb S}^2$ whose horizontal lift is an open trajectory in $SU(2)$ \cite{LHC}.} The ground state is invariant to actions of the maximal compact subgroup. In our case, $U(1) \cong SO(2)$ is the group of rotations in the complex plane, and the uniform measure on $S^1$ is invariant w.r. to actions of this group. Hence, the ground state is represented by uniform distribution. All coherent states are obtained by actions of $SU(1,1)$ on the ground state, which yields precisely the Poisson manifold. The quantum evolution is described in terms of parameters of $SU(1,1)$.

In whole, the group-theoretic study of TW's unveils the same geometric setup as one arising in the cyclic evolution of $SU(1,1)$-coherent states.

\subsection{Geometric phase as a hidden variable for evolution on coherent states}
\label{sec:7_2}
We have pointed out that geometric phase appears as a hidden variable in models (\ref{chimera}) and (\ref{conf-contr}). It is invisible on the macroscopic level (when observing densities), but its impact is transparent on the microscopic level (when observing individual oscillators). This entails a counter-intuitive phenomenon that rather complicated (deterministic) individual motions coalesce into a stationary (up to rotations) density.

We conclude this Section by explaining that the geometric phase appears as a hidden variable only when the dynamics are restricted to the manifold of coherent states.

We say that a probability measure on the circle is {\it balanced}, if its mean value (mathematical expectation) is zero.

\begin{theorem}
Let $\mu$ be an absolutely continuous unbalanced probability measure on $S^1$. Suppose that $\mu$ is invariant under some transformation $g \in {\mathbb G}$ that is not identity. Then, $\mu$ is a measure whose density is the Poisson kernel.
\end{theorem}

The proof is based on the notion of conformal barycenter (and the fact that Poisson kernels are the only probability measures for which conformal barycenter coincides with the mean value). We omit this proof, as it can be found in our previous paper \cite{CJ-IJMP-B}.

\begin{corollary}
Let $\rho_A(\varphi)$ and $\rho_B(\varphi)$ be two Poisson kernels with parameters $\alpha_A = r_A e^{i \psi_A}$ and $\alpha_B = r_B e^{i \psi_B}$, respectively. Then, any transformation $g \in {\mathbb G}$, satisfying $g(\alpha_A) = \alpha_B$ maps the measure whose density is $\rho_A$ into the measure with density $\rho_B$.
\end{corollary}
\begin{figure*}[t]
\centering
  \begin{tabular}{@{}ccc@{}}
    \includegraphics[width=.33\textwidth]{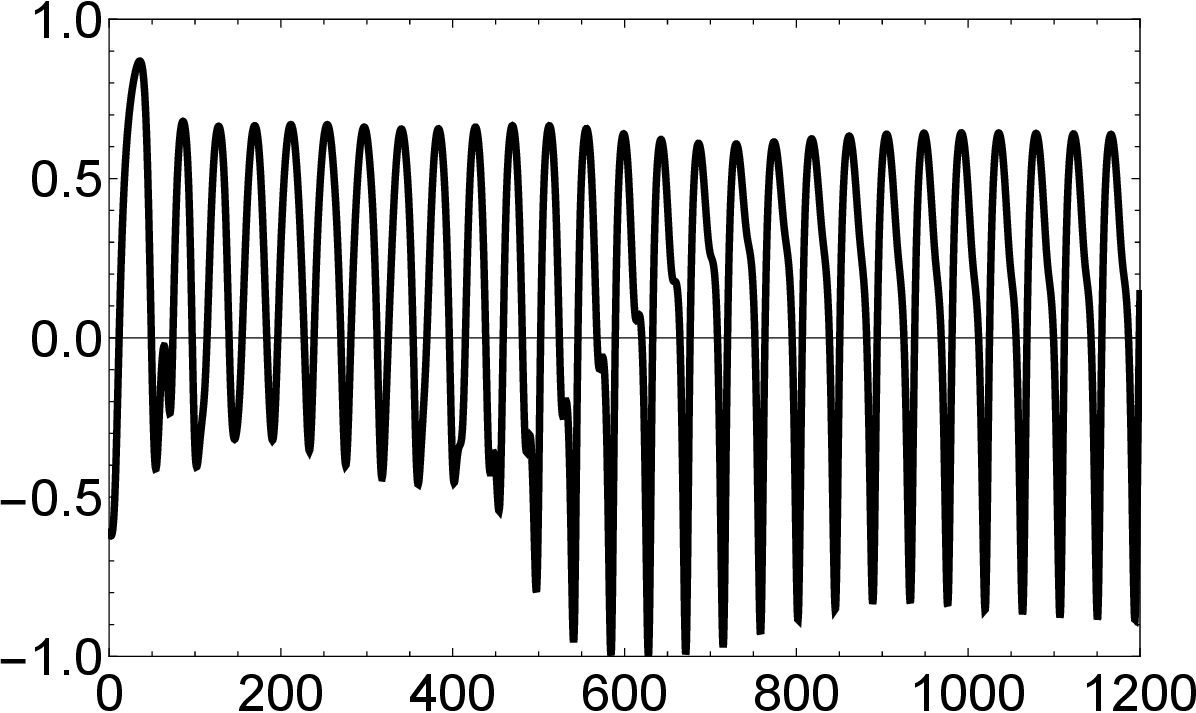} &
    \includegraphics[width=.33\textwidth]{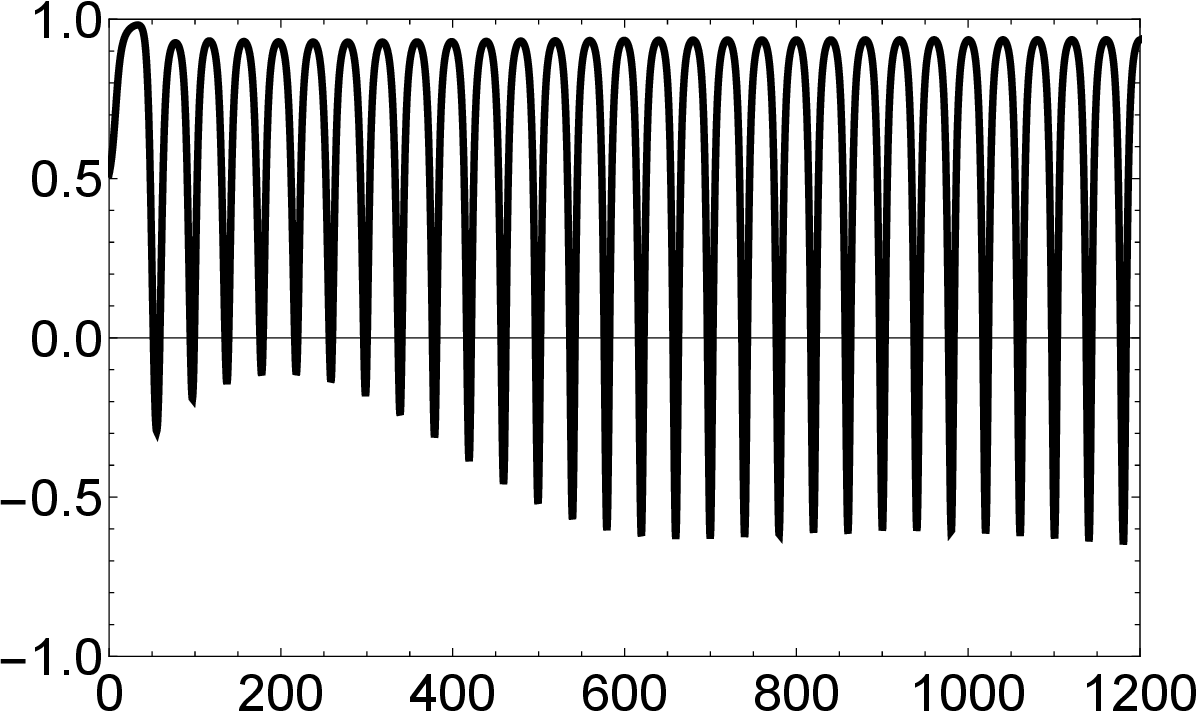}&
    \includegraphics[width=.33\textwidth]{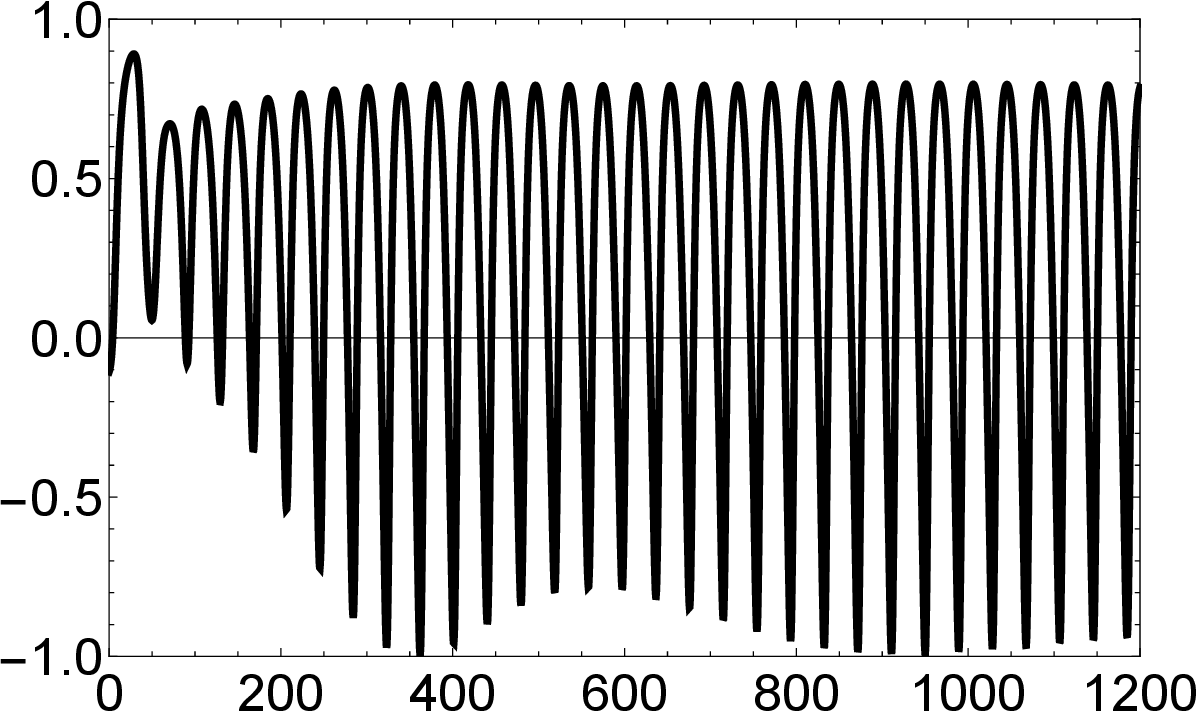}\\
     \quad a)&\quad b)&\quad c)
  \end{tabular}
  \caption{\label{fig:8} Evolution of scalar products between two random contrarians as the system (\ref{conf-contr-sphere}) evolves towards the traveling wave state for a model on (a) $S^2$; (b) $S^3$ and (c) $S^4$. Obtained for $N=500$ oscillators, with $p=0.7$, $K_E =0.7$, $K_F =-0.3$.}
\end{figure*}

The above Corollary states that if two Poisson kernels have equal real order parameters (i.e. $r_A = r_B$), then the rotation is not unique M\" obius transformation that maps one into another.
The above Theorem exposes that the Poisson submanifold is the unique orbit of ${\mathbb G}$ with such a property.

\section{Holonomy of non-Abelian phase in conf-contr models on spheres}
\label{sec:8}
In sections \ref{sec:4} and \ref{sec:5} we investigated a holonomy of parallel transport in the fiber bundle $SU(1,1)$ for models (\ref{chimera}) and (\ref{conf-contr}). The fibers are circles, i.e. elements of the Abelian group $U(1) \cong SO(2)$.

Along with Abelian geometric phases, the notion of geometric phase extends to physical systems evolving on fiber bundles with non-Abelian fibers \cite{Anandan, Levay}.

Recently, the authors considered an extension of the conf-contr model to spheres and proven that TW's on spheres arise in such models \cite{CJN}. In the present Section we point out that these TW's on spheres are associated with holonomies along the non-Abelian group $SO(d)$.

We first introduce the conf-contr models on spheres. Let $x_j^l, \, l=E,F$ be unit vectors in the $d$-dimensional vector space, representing generalized oscillators and let $W$ be an anti-symmetric matrix, interpreted as a (generalized) frequency. Note that we assume that oscillators are identical.

The conf-contr model on the $d-1$-dimensional sphere $S^{d-1}$ reads \cite{CJN}
\begin{equation}
\label{conf-contr-sphere}
\begin{cases}
\dot x_j^E = W x_j^E + f_E - \langle x_j^E,f_E \rangle x_j^E, \quad j=\overline{1,M}; \\
\dot x_j^F = W x_j^F + f_F - \langle x_j^F,f_F \rangle x_j^F, \quad j=\overline{1,N},
\end{cases}\,
\end{equation}
where the notion $\langle \cdot,\cdot \rangle$ stands for the inner product in ${\mathbb R}^n.$ Further, $f_l = f_l(x_1,\dots,x_N), \, l=E,F$ are vector-valued coupling functions given by
\begin{equation}
\label{coupling}
\begin{split}
f_E &= \frac{K_E}{M+N} \left(\sum \limits_{i=1}^M x_i^E + \sum \limits_{i=1}^N x_i^F \right),\\
f_F &= \frac{K_F}{M+N} \left( \sum \limits_{i=1}^M x_i^E + \sum \limits_{i=1}^N x_i^F \right),
\end{split}
\end{equation}
where $K_E > 0, \; K_F<0$.
This means that oscillators $x_j^E$ ("conformists") are positively coupled to the mean field, while $x_j^F$ ("contrarians") are negatively coupled.
Again, introduce parameters $p = M/(M+N)$ and $q=N/(M+N)$ representing portions of conformists and contrarians in the whole population, respectively.

The dynamics (\ref{conf-contr-sphere}) can be reduced to the low-dimensional invariant submanifold. This reduction has been exploited \cite{CJN} in order to conduct stability analysis of equilibrium states in (\ref{conf-contr-sphere}). It has been shown that for certain values of parameters $p$ and $K_E / K_F$ there exists a stable equilibrium in which conformists achieve full synchronization, while contrarians are only partially synchronized. The equilibrium values of the real order parameters are $r_E = 1; \, r_F = const < 1$. In addition, the two sub-populations evolve on ${\mathbb S}^{d-1}$ preserving the constant angle $\delta \in (\pi/2,\pi)$ between them. Hence, contrarians are organized into the TW on a sphere.

In order to conduct a group-theoretic study of TW's on spheres, consider the group ${\mathbb M}$ of transformations acting on the $d$-dimensional unit ball ${\mathbb B}^d$ by the following formula
$$
m(x) = R \left( \frac{(-x + |x|^2a)(1-|a|^2)}{1 - \langle a,x \rangle + |a|^2 |x|^2} + a \right),
$$
where $a \in {\mathbb B}^d$, $R \in SO(d)$.

${\mathbb M}$ is the group of isometries of the unit ball in hyperbolic metric. It is isomorphic to the Lorentz group $SO^+(d,1)$. Generalized oscillators in (\ref{conf-contr-sphere}) evolve by actions of ${\mathbb M}$ \cite{LMS}. A slight adaptation of this result yields the following two propositions.

\begin{proposition}
\label{prop:6}
There exist two one-parametric families of transformations $m_t^E,m_t^F \in {\mathbb M}$, such that:
$$
x_j^E(t) = m_j^E(x_j^E(0)) \mbox{  and  } x_j^F(t) = m_j^F(x_j^F(0)),
$$
$\forall j=1,\dots,N$ and $\forall t>0$.
\end{proposition}
\begin{figure*}[t]
\centering
  \begin{tabular}{@{}ccc@{}}
    \includegraphics[width=.27\textwidth]{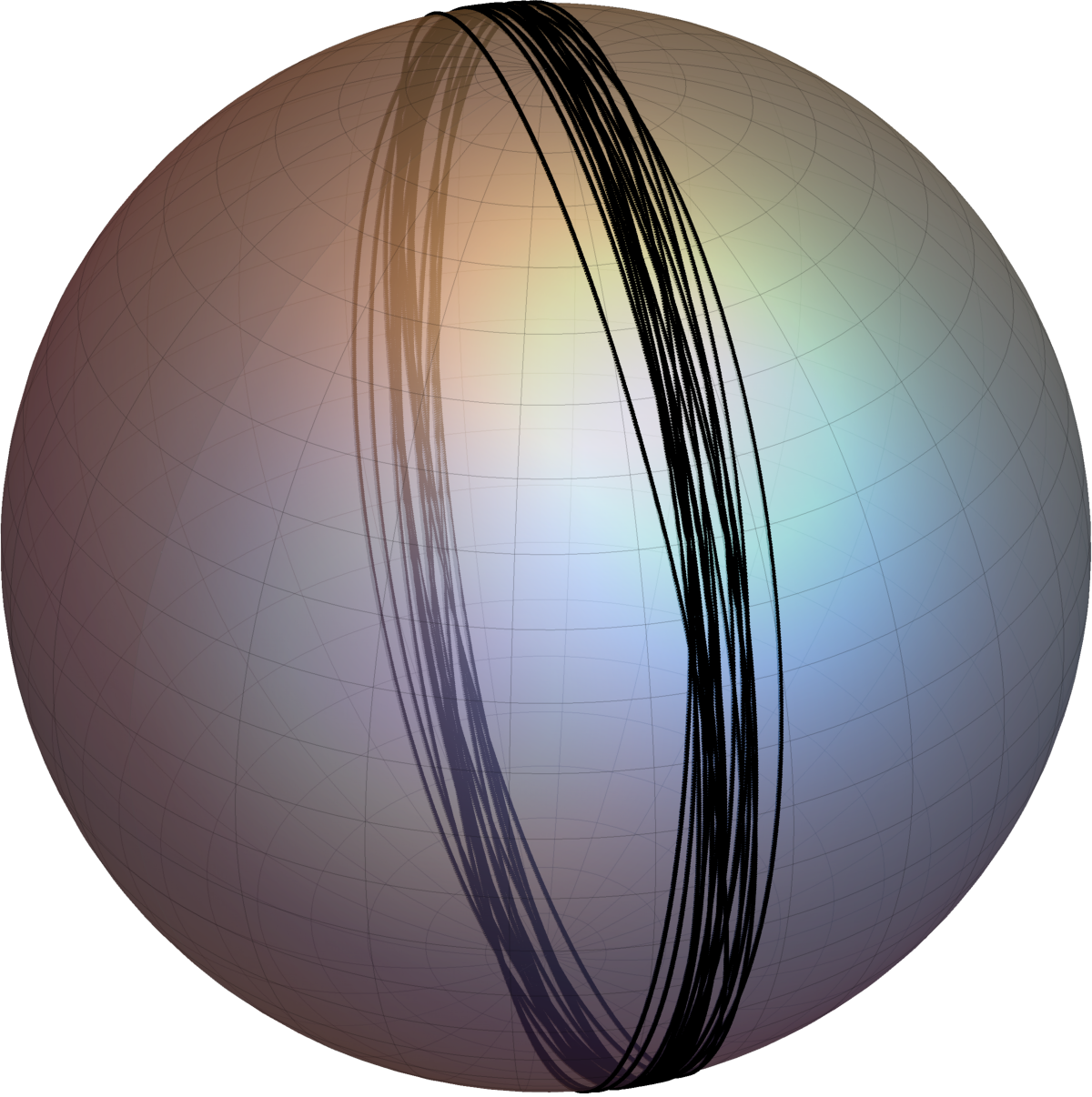} &
    \includegraphics[width=.27\textwidth]{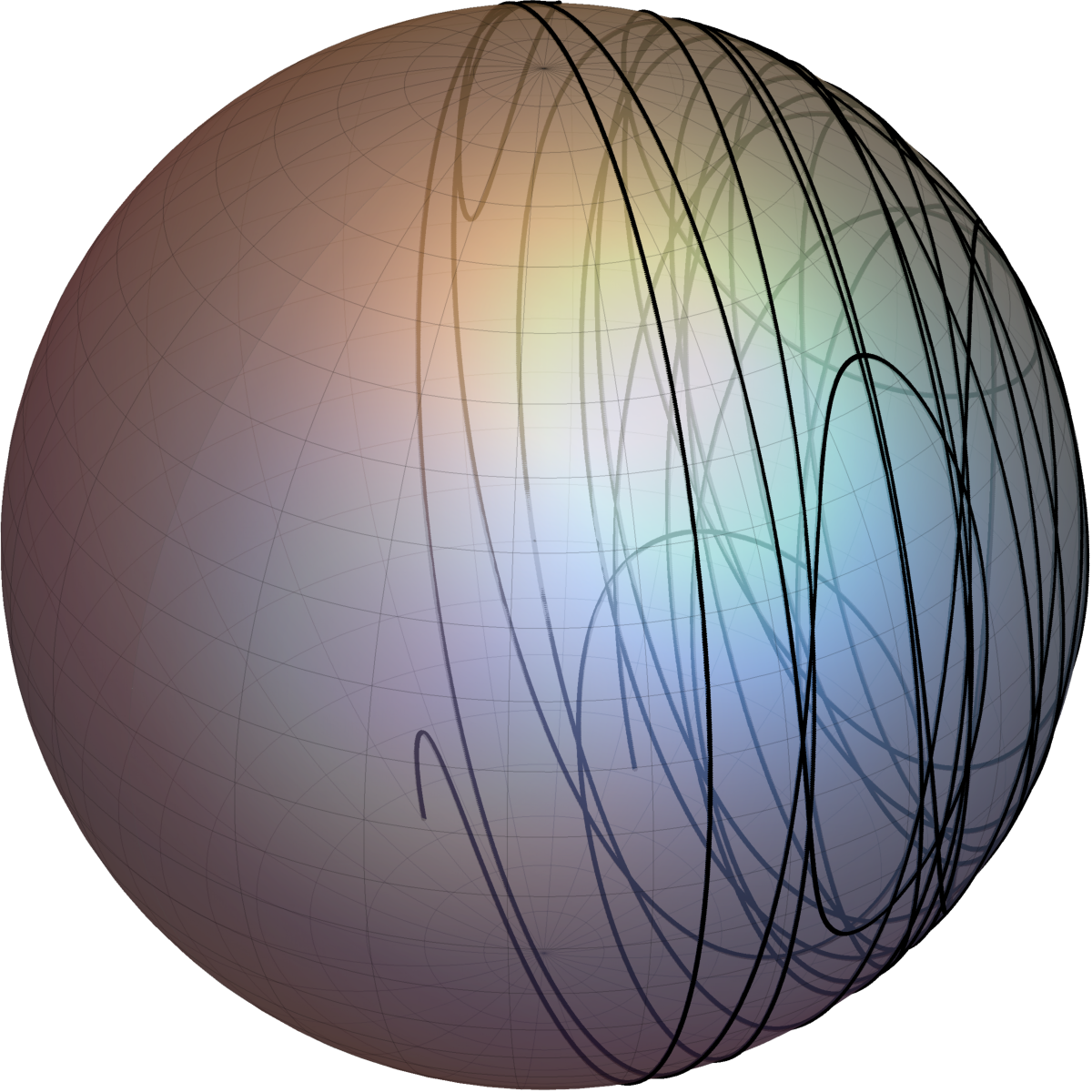}&
    \includegraphics[width=.27\textwidth]{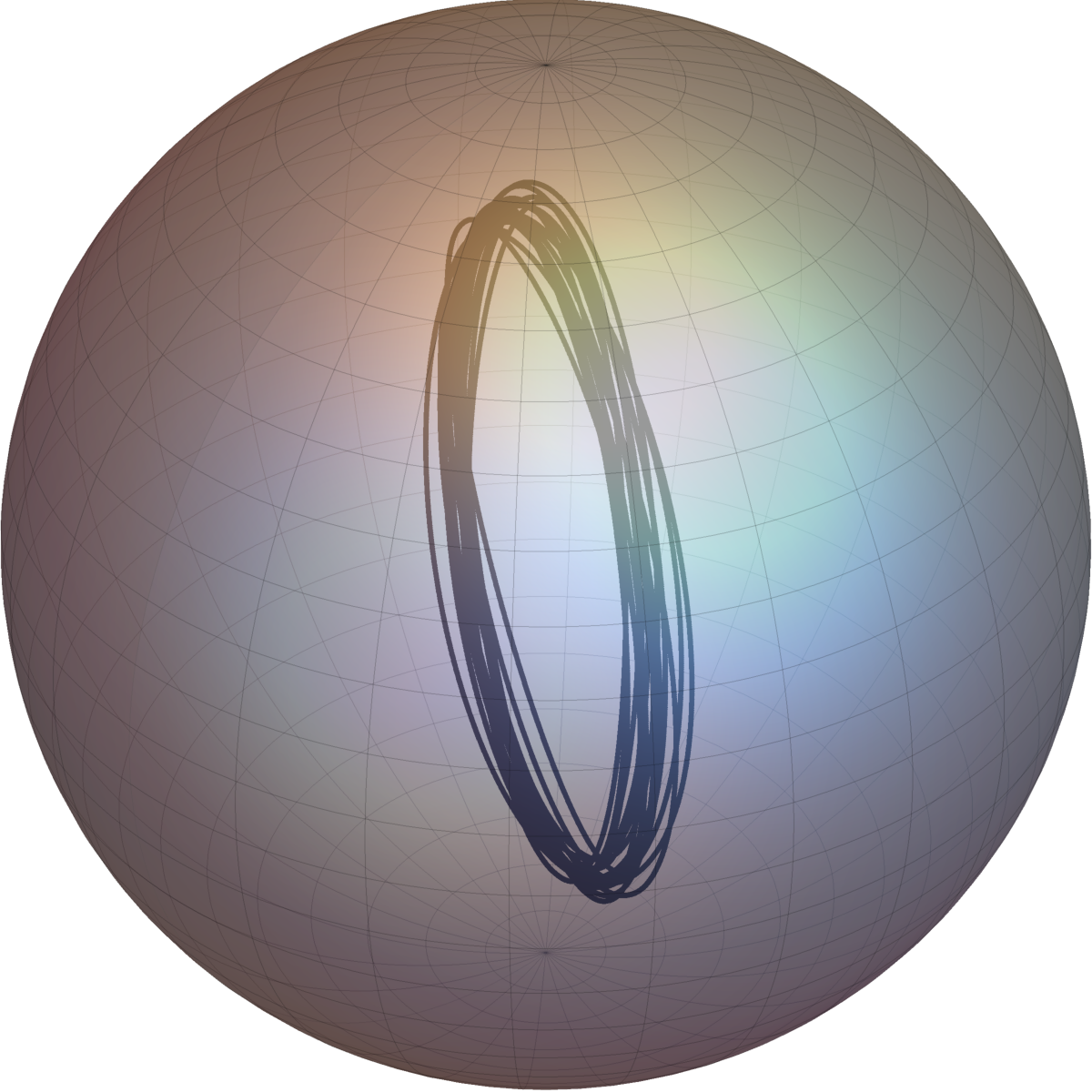}\\
     \quad a)&\quad b)&\quad c)
  \end{tabular}
  \caption{\label{fig:9} Trajectories in the model (\ref{conf-contr-sphere}) on $S^2$ in the traveling wave state (for the same parameter values as in Figure \ref{fig:8}: (a) conformists; (b) a random contrarian and (c) centroid of contrarians.}
\end{figure*}
\begin{proposition}
\label{prop:7}
Consider the system (\ref{conf-contr-sphere}) in thermodynamic limit $N \to \infty$. Assume that initial distributions of conformists and contrarians are uniform on ${\mathbb S}^{d-1}$. Then the densities of conformists and contrarians at each moment $t$ are of the following form
\begin{equation}
\label{hyperbolic_Poisson_kernel}
\rho_{hyp}(x) = \frac{\Gamma(d/2)}{2 \pi^{d/2}} \left( \frac{1-|a_l|^2}{|a_l-x|^2} \right)^{d-1}, \quad l=E,F,
\end{equation}
where $a_l \in {\mathbb B}^d$ and $x \in {\mathbb S}^{d-1}$.
\end{proposition}

\begin{definition}
Submanifold of densities of the form (\ref{hyperbolic_Poisson_kernel}) is named the $d$-dimensional Poisson manifold.
\end{definition}

This submanifold can be identified with the hyperbolic unit ball ${\mathbb B}^d$ and is invariant for actions of ${\mathbb M}$. Uniform and delta distributions on ${\mathbb S}^{d-1}$ belong to the $d$-dimensional cases, as limit cases for $a_l = 0$ and $|a_l| \to 1$, respectively.

Due to Proposition \ref{prop:6}, sub-population of contrarians in the model  (\ref{conf-contr-sphere}) generates a trajectory in ${\mathbb M}$. Moreover, from Proposition \ref{prop:7} it follows that if initial distributions are uniform, the dynamics are restricted on the $d$-dimensional Poisson manifold, or, equivalently inside the ball ${\mathbb B}^d$. Since the real order parameter $r_F$ is constant, this trajectory is confined on the sphere with the radius $r_F < 1$ inside ${\mathbb B}^d$.

The horizontal lift of this trajectory is a trajectory on the full state space ${\mathbb M}$ with the dimension $d(d-1) / 2 + d$. Fibers of this projection from ${\mathbb M}$ to the sphere inside ${\mathbb B}^d$ are elements of the special orthogonal group $SO(d)$.

Evolutions of scalar products between random contrarians on spheres ${\mathbb S}^2,{\mathbb S}^3,{\mathbb S}^4$ are plotted in Figure \ref{fig:8}. These simulations imply that individual oscillators evolve by transformations from ${\mathbb M}$ that are not rotations in the $d$-dimensional vector space.

In Figure \ref{fig:9} we plot trajectories (\ref{conf-contr-sphere}) on the sphere ${\mathbb S}^2$. Although it looks like the density performs $SO(d)$ rotations (Figure \ref{fig:9}c), the dynamics of individual contrarians (Figure \ref{fig:9}b) turn out to be more complicated.
This unveils an impact of the non-Abelian geometric $SO(d)$-phase on dynamics. Notice that evolution of densities (\ref{hyperbolic_Poisson_kernel}) in this case is not cyclic, as TW rotates, but does not necessarily trace a closed loop on ${\mathbb S}^{d-1}$.

\section{Conclusion}
\label{sec:9}
The present study demonstrates that intriguing temporal patterns in two-populations Kuramoto models can be understood through the group-theoretic and geometric approach. We have shown that these unexpected equilibrium states are associated with the phase holonomy. Indeed, suppose that we vary parameters in models (\ref{chimera}), (\ref{conf-contr}) or (\ref{conf-contr-sphere}). As the parameters pass a certain critical boundary, simple stationary equilibria bifurcate into TW's. Emergence of the TW is associated with the birth of phase holonomy in all three models. This unveils a geometric subtlety standing behind unexpected dynamical behaviors in simple networks. This also provides a geometric explanation of the recently reported fact that such equilibria are at most neutrally stable in the full state space \cite{EM}.

From the mathematical point of view, holonomy can be regarded as a manifestation of the geometric phase. Geometric phase is a common term referring to various effects observed in many fields of Physics. Although seemingly unrelated, all these manifestations are explained through the mathematical concept of holonomy of parallel translation for a connection on the fiber bundle. In quantum systems, the geometric phase is related to the fibration $SU(2)/U(1) = {\mathbb S}^2$ corresponding to the quantum spin. In Kuramoto models, the holonomy groups are Lorentz groups: $SU(1,1) \cong SO^+(2,1)$ in (\ref{chimera}) and (\ref{conf-contr}) and $SO^+(d,1)$ in (\ref{conf-contr-sphere}).

In the case of TW's on spheres arising in (\ref{conf-contr-sphere}), we report the non-Abelian geometric phase, along the $SO(d)$ fibers of the Lorentz group $SO^+(d,1)$.

Our findings point out that the motions of TW's arising in two-populations Kuramoto models are deceptively simple. TW as a whole appears to evolve by simple rotations, but individual motions are more complicated.

Although references on geometric phases for coherent states are numerous, particular manifestations with the holonomy group $SU(1,1)$ are not frequently reported. A nice example of this kind is the Prytz planimeter, one of classical devices invented in XIX century for measuring areas. The geometric mechanism behind the Prytz planimeter has been explained as parallel translation for a connection on the circle bundle acted on by $SU(1,1)$ \cite{Foote}. We find it remarkable that subtleties of TW's in ensembles of Kuramoto oscillators can be explained by the same geometric phenomenon as an old device for measuring areas on geographic maps.



\nocite{*}
\bibliography{aipsamp}

\end{document}